\documentclass[useAMS]{mn2e}
\usepackage{times}
\usepackage{psfig}
\usepackage{amssymb}
\usepackage{amsbsy}
\usepackage{color}

\def\as0{6.82 \pm 0.28}
\def\agams{1.57 \pm 0.07}
\def\ar0{5.05 \pm 0.26}
\def\agamr{1.67 \pm 0.03}
\def\betaQ{0.47^{+0.19}_{-0.16}}
\def\betaxirat{0.45 \pm 0.14}
\def\xirat{1.34 \pm 0.13}
\def\Qval{0.55 \pm 0.18}
\def\betaoverall{0.49 \pm 0.09}
\def\aoverall{506 \pm 52}

\def\bj{b_{\rm J}}
\def\sqdeg{{\rm~deg}^2}
\def\xisp{\xi(\sigma, \pi)}
\def\hmpc{\;h^{-1}{\rm Mpc}} 
 
\def\hmpcb{h^{-1}{\rm Mpc}}
\def\kms{\rm{\;km\;s^{-1}}}
\def\kmsn{\rm{km\;s^{-1}}}
\def\kmsmpc{\rm{\;km\;s^{-1}Mpc^{-1}}} 
\def\ct{\mu} 
\def\etal{\rm{et al.}}
\def\eg{e.g.~} 
\def\refer{\par \noindent \hangindent=0.4cm \hangafter=1}
\def\pcf{\Xi(\sigma)/\sigma} 
\def\oml{\Omega_{\Lambda}}
\def\om{\Omega_{\rm{m}}} 
\def\lcdm{\Lambda{\rm{CDM}}}

\def\xis{\xi(s)}
\def\xir{\xi(r)}
\def\nts{\negthinspace}

\begin{document}

\title[The 2dFGRS: correlation functions, peculiar velocities and the
matter density of the Universe]{\vspace*{-0.5cm}The 2dF Galaxy
Redshift Survey: correlation functions, peculiar velocities and the
matter density of the Universe} \author[Hawkins et al. (The 2dFGRS
Team)]{
\parbox[t]{\textwidth}{\vspace*{-1.0cm}Ed Hawkins$^1$\footnotemark[1],
Steve Maddox$^1$\footnotemark[1], Shaun Cole$^2$, Ofer Lahav$^3$, 
Darren S.\ Madgwick$^{3,4}$, Peder Norberg$^{2,5}$, John A.\ 
Peacock$^6$, Ivan K. Baldry$^7$, Carlton M.\ Baugh$^2$, Joss 
Bland-Hawthorn$^8$, Terry Bridges$^8$, Russell Cannon$^8$, Matthew 
Colless$^9$, Chris Collins$^{10}$, Warrick Couch$^{11}$, Gavin 
Dalton$^{12, 13}$, Roberto De Propris$^{11}$, Simon P.\ Driver$^9$, 
George Efstathiou$^3$, Richard S.\ Ellis$^{14}$, Carlos S.\ Frenk$^2$, 
Karl Glazebrook$^7$, Carole Jackson$^9$, Bryn Jones$^1$, Ian Lewis$^{12}$,
Stuart Lumsden$^{15}$, Will Percival$^6$, Bruce A.\ Peterson$^9$, Will
Sutherland$^6$ and Keith Taylor$^{14}$ (The 2dFGRS Team)\\ 
\small{\it
$^1$School of Physics \& Astronomy, University of Nottingham,
Nottingham NG7 2RD, UK\\ 
$^2$Department of Physics, University of
Durham, South Road, Durham DH1 3LE, UK \\ 
$^3$Institute of Astronomy,
University of Cambridge, Madingley Road, Cambridge CB3 0HA, UK \\
$^4$Department of Astronomy, University of California at Berkeley,
Berkeley, CA 94720, USA\\ 
$^5$Institut f\"ur Astronomie, ETH H\"onggerberg, CH-8093 Z\"urich,
Switzerland\\
$^6$Institute for Astronomy, University of Edinburgh, Royal Observatory, Blackford Hill, Edinburgh EH9 3HJ, UK\\ 
$^7$Department of Physics \& Astronomy, Johns Hopkins University,
Baltimore, MD 21218-2686, USA \\ 
$^8$Anglo-Australian Observatory,
P.O.\ Box 296, Epping, NSW 2121, Australia\\ 
$^9$Research School of
Astronomy \& Astrophysics, The Australian National University, Weston
Creek, ACT 2611, Australia \\ 
$^{10}$Astrophysics Research Institute, Liverpool John Moores University, Twelve Quays House, Birkenhead, L14 1LD, UK \\ 
$^{11}$Department of Astrophysics, University of New South
Wales, Sydney, NSW 2052, Australia \\ 
$^{12}$Department of Physics,
University of Oxford, Keble Road, Oxford OX1 3RH, UK \\ 
$^{13}$Space Science and Technology Division, Rutherford Appleton Laboratory,
Chilton, Didcot, OX11 0QX, UK \\ 
$^{14}$Department of Astronomy,
California Institute of Technology, Pasadena, CA 91125, USA \\
$^{15}$Department of Physics, University of Leeds, Woodhouse Lane,
Leeds, LS2 9JT, UK\\}
\footnotemark[1]~E-mail: ppxeh@nottingham.ac.uk (EH), steve.maddox@nottingham.ac.uk (SM) 
}}

\date{Recieved 2003 May 8; in original form 2002 December 15.}
\pagerange{\pageref{firstpage}--\pageref{lastpage}} 
\pubyear{2003}

\label{firstpage}

\maketitle

\begin{abstract}
We present a detailed analysis of the two-point correlation function,
$\xisp$, from the 2dF Galaxy Redshift Survey (2dFGRS). The large size
of the catalogue, which contains $\sim 220\,000$ redshifts, allows us
to make high precision measurements of various properties of the
galaxy clustering pattern. The effective redshift at which our
estimates are made is $z_s \approx 0.15$, and similarly the effective
luminosity, $L_s \approx 1.4L^{\ast}$. We estimate the redshift-space
correlation function, $\xis$, from which we measure the redshift-space
clustering length, $s_0 = \as0\hmpc$. We also estimate the projected
correlation function, $\Xi(\sigma)$, and the real-space correlation
function, $\xir$, which can be fit by a power-law
$(r/r_0)^{-\gamma_r}$, with $r_0 = \ar0\hmpc$, $\gamma_r =
\agamr$. For $r \gtrsim 20\hmpc$, $\xi$ drops below a power-law as,
for instance, is expected in the popular $\lcdm$ model. The ratio of
amplitudes of the real and redshift-space correlation functions on
scales of $8 - 30\hmpc$ gives an estimate of the redshift-space
distortion parameter $\beta$. The quadrupole moment of $\xisp$ on
scales $30 - 40\hmpc$ provides another estimate of $\beta$. We also
estimate the distribution function of pairwise peculiar velocities,
$f(v)$, including rigorously the significant effect due to the infall
velocities, and find that the distribution is well fit by an
exponential form. The accuracy of our $\xisp$ measurement is
sufficient to constrain a model, which simultaneously fits the shape
and amplitude of $\xir$ and the two redshift-space distortion effects
parameterized by $\beta$ and velocity dispersion, $a$. We find $\beta
= \betaoverall$ and $a = \aoverall\kms$, though the best fit values
are strongly correlated. We measure the variation of the peculiar
velocity dispersion with projected separation, $a(\sigma)$, and find
that the shape is consistent with models and simulations. This is the
first time that $\beta$ and $f(v)$ have been estimated from a
self-consistent model of galaxy velocities. Using the constraints on
bias from recent estimates, and taking account of redshift evolution,
we conclude that $\beta(L = L^*, z = 0) = 0.47 \pm 0.08$, and that the
present day matter density of the Universe, $\om \approx 0.3$,
consistent with other 2dFGRS estimates and independent analyses.
\end{abstract}


\section{Introduction}

The galaxy two-point correlation function, $\xi$, is a fundamental
statistic of the galaxy distribution, and is relatively
straightforward to calculate from observational data. Since the
clustering of galaxies is determined by the initial mass fluctuations
and their evolution, measurements of $\xi$ set constraints on the
initial mass fluctuations and their evolution.  The astrophysics of
galaxy formation introduces uncertainties, but there is now good
evidence that galaxies do trace the underlying mass distribution on
large scales.

In this paper we analyse the distribution of $\sim$ 220\,000 galaxies
in the 2-degree Field Galaxy Redshift Survey (2dFGRS, Colless
\etal~2001). A brief summary of the data is presented in Section
2. Much of our error analysis makes use of mock galaxy catalogues
generated from $N$-body simulations which are also discussed in
Section 2.

The two-dimensional measurement $\xisp$, where $\sigma$ is the pair
separation perpendicular to the line-of-sight and $\pi$ is the pair
separation parallel to the line-of-sight provides information about
the real-space correlation function, the small-scale velocity
distribution, and the systematic gravitational infall into overdense
regions. The spherical average of $\xisp$ gives an estimate of the
redshift-space correlation function, $\xis$, where $s=\sqrt{(\pi^2 +
\sigma^2)}$, so the galaxy separations are calculated assuming that
redshift gives a direct measure of distance, ignoring the effects of
peculiar velocities. Integrating $\xisp$ along the line-of-sight sums
over any peculiar velocity distributions, and so is unaffected by any
redshift-space effects. The resulting projected correlation function,
$\Xi(\sigma)$, is directly related to the real-space correlation
function.  Our estimates of $\xisp$, $\xis$, $\Xi(\sigma)$ and $\xir$
are presented in Section 3. These statistics have been measured from
many smaller redshift surveys (\eg Davis \& Peebles 1983; Loveday
\etal~1992; Jing, Mo \& B\"orner~1998; Hawkins \etal~2001; Zehavi
\etal~2002), but since they sample smaller volumes, there is a large
cosmic variance on the results. The large volume sampled by the 2dFGRS
leads to significantly more reliable estimates. A preliminary analysis
was performed on the 2dFGRS by Peacock \etal~(2001) but we now have a
far more uniform sample and twice as many galaxies. Madgwick
\etal~(2003) have measured these statistics for spectral-type
sub-samples of the 2dFGRS.

Peculiar velocities of galaxies lead to systematic differences between
redshift-space and real-space measurements, and we can consider the
effects in terms of a combination of large-scale coherent flows
induced by the gravity of large-scale structures, and a small-scale
random peculiar velocity of each galaxy (e.g. Marzke \etal~1995; Jing
\etal~1998). The large-scale flows compress the contours of $\xisp$
along the $\pi$ direction, as described by Kaiser (1987) and Hamilton
(1992). The amplitude of the distortion depends on the mean density of
the universe, $\om$, and on how the mass distribution is clustered
relative to galaxies, which can be parameterized in terms of a linear
bias $b$, defined so that $\delta_g = b\delta_m$, where $\delta$
represents fluctuations in the density field. The random component of
peculiar velocity for each galaxy means that the observed $\xisp$ is
convolved in the $\pi$ coordinate with the pairwise distribution of
random velocities. In section 4 we describe the construction of a
model $\xisp$ from these assumptions about redshift-space distortions
and also the shape of the correlation function.

In section 5 we use the $Q$ statistic (Hamilton 1992) based on the
quadrupole moment of $\xisp$ to estimate the parameter
$\beta\approx\om^{0.6}/b$. In the absence of the small-scale random
velocities the shape of $\xisp$ contours on large scales is directly
related to the parameter $\beta$. A similar estimate of $\beta$ is
provided by the ratio of amplitudes of $\xis$ to $\xir$ and this
is also presented in Section 5.

In section 6, we use the Landy, Szalay \& Broadhurst (1998) method to
estimate the distribution of peculiar velocities. This technique
ignores the effect of large-scale distortions and uses the Fourier
transform of $\xisp$ to estimate the distribution of peculiar
velocities, $f(v)$.  The large sample volume of the 2dFGRS makes our
measurements more reliable than previous estimates in the same way as
for the correlation functions mentioned earlier.

These two approaches provide reasonable estimates of $\beta$ and
$f(v)$ so long as the distortions at small and large scales are
completely decoupled. This is not the case for real data, and so we
have fitted models which simultaneously include the effects of both
$\beta$ and $f(v)$. The resulting best-fit parameters are the most
self consistent estimates. Previous data-sets have lacked the
signal-to-noise to allow a reliable multi-parameter fit in this
way. Our fitting procedure and results are described in Section 7.

In Section 8, we examine the luminosity and redshift dependence of
$\beta$ and combine our results with estimates of $b$ (Verde
\etal~2002; Lahav \etal~2002) to estimate $\om$, and compare this with
other recent analyses.

In Section 9, we summarise our main conclusions. When converting from
redshift to distance we assume the Universe has a flat geometry with
$\oml = 0.7$, $\om = 0.3$ and $H_0 = 100\,h\kmsmpc$, so that all scales
are in units of $\hmpc$.

\section{The Data}

\label{s:data}

\subsection{The 2dFGRS data}
\label{s:2dfdata}

The 2dFGRS is selected in the photometric $\bj$\ band from the APM
galaxy survey (Maddox, Efstathiou \& Sutherland 1990) and its
subsequent extensions (Maddox \etal, in preparation). The bulk of the
solid angle of the survey is made up of two broad strips, one in the
South Galactic Pole region (SGP) covering approximately
$-37^\circ\negthinspace.5<\delta<-22^\circ\negthinspace.5$, $21^{\rm
h}40^{\rm m}<\alpha<3^{\rm h}40^{\rm m}$ and the other in the
direction of the North Galactic Pole (NGP), spanning
$-7^\circ\negthinspace.5<\delta<2^\circ\negthinspace.5$, $9^{\rm
h}50^{\rm m}<\alpha<14^{\rm h}50^{\rm m}$.  In addition to these
contiguous regions, there are a number of circular 2-degree fields
scattered randomly over the full extent of the low extinction regions
of the southern APM galaxy survey.

The magnitude limit at the start of the survey was set at $\bj =
19.45$ but both the photometry of the input catalogue and the dust
extinction map have been revised since and so there are small
variations in magnitude limit as a function of position over the
sky. The effective median magnitude limit, over the area of the
survey, is $\bj \approx 19.3$ (Colless \etal~2001).

\begin{figure*}
\psfig{figure=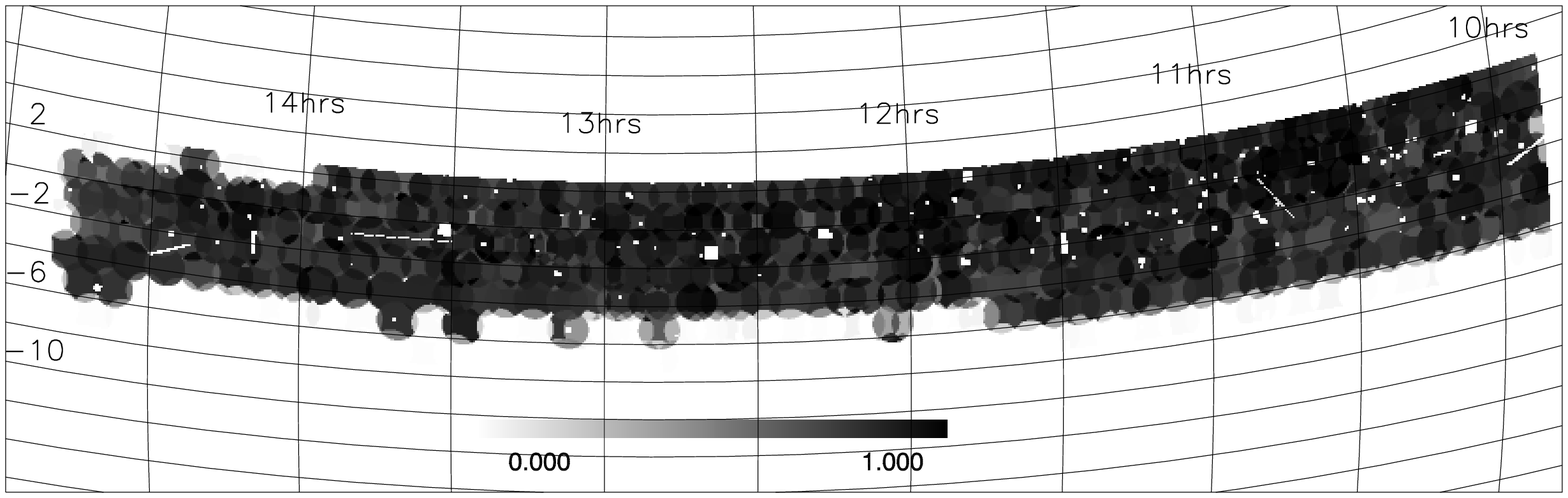,angle=0,width=\textwidth,clip=}
\psfig{figure=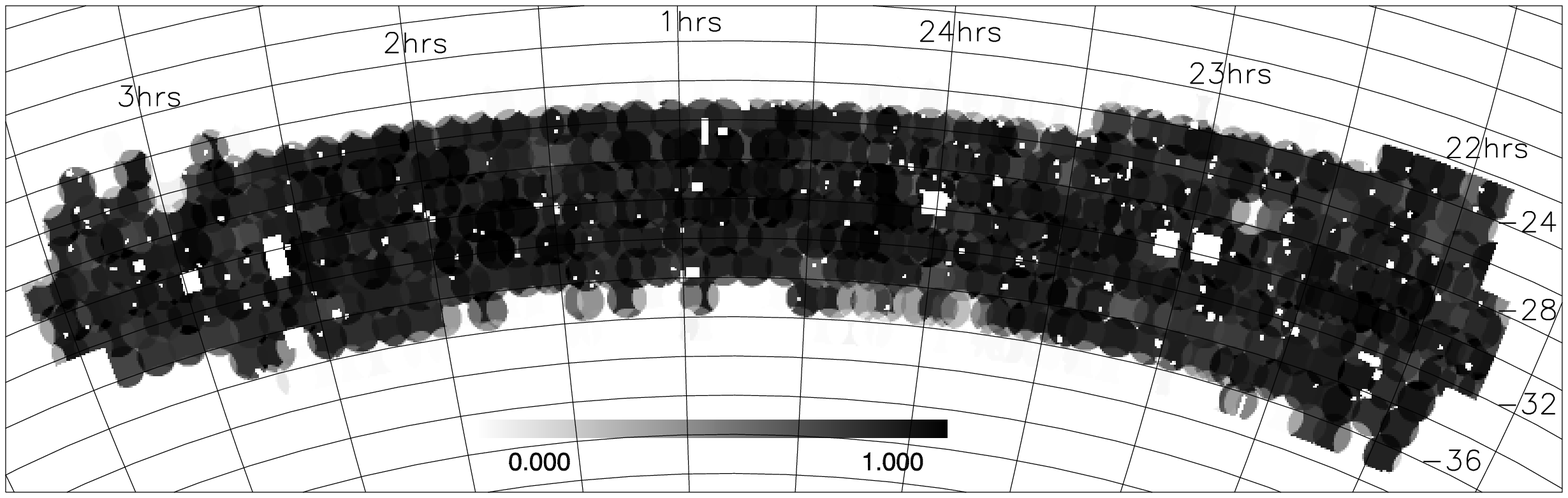,angle=0,width=\textwidth,clip=}
\caption[]{The redshift completeness masks for the NGP (top) and SGP
(bottom). The greyscale shows the completeness fraction.}
\label{f:masks}
\end{figure*}

The completeness of the survey data varies according to the position
on the sky because of unobserved fields (mostly around the survey
edges), un-fibred objects in observed fields (due to collision
constraints or broken fibres) and observed objects with poor
spectra. The variation in completeness is mapped out using a
completeness mask (Colless \etal~2001; Norberg \etal~2002a) which is
shown in Fig.\ \ref{f:masks} for the data used in this paper.

We use the data obtained prior to May 2002, which is virtually the
completed survey. This includes 221\,283 unique, reliable galaxy
redshifts (quality flag $\geqslant$ 3, Colless \etal~2001). We analyse
a magnitude-limited sample with redshift limits $z_{\rm{min}} = 0.01$
and $z_{\rm{max}} = 0.20$, and no redshifts are used from a field with
$< 70\%$ completeness. The median redshift is $z_{\rm med} \approx
0.11$. The random fields, which contain nearly 25\,000 reliable
redshifts are not included in this analysis. After the cuts for
redshift, completeness and quality we are left with 165\,659 galaxies
in total, 95\,929 in the SGP and 69\,730 in the NGP. These data cover
an area, weighted by the completeness shown in Fig.\ \ref{f:masks}, of
$647\sqdeg$ in the SGP and $446\sqdeg$ in the NGP, to the magnitude
limit of the survey.

In all of the following analysis we consider the NGP and SGP as
independent data sets. Treating the NGP and SGP as two independent
regions of the sky gives two estimates for each statistic, and so
provides a good test of the error bars we derive from mock catalogues
(see below). We have also combined the two measurements to produce our
overall best estimate by simply adding the pair counts from the NGP
and SGP. The optimal weighting of the two estimates depends on the
relative volumes surveyed in the NGP and SGP, but since these are
comparable, a simple sum is close to the optimal combination.

It is important to estimate the effective redshift at which all our
statistics are calculated. As $\xi$ is based on counting pairs of
galaxies the effective redshift is not the median, but a pair-weighted
measure. The tail of high redshift galaxies pushes this effective
redshift to $z_s \approx 0.15$. Similarly the effective magnitude of
the sample we analyse is $M_s - 5\log h \approx -20.0$, corresponding
to $L_s \approx 1.4L^*$ (using $M^{\ast} - 5\log h = -19.66$, Norberg
\etal~2002a).

\subsection{Mock catalogues}
\label{s:mocks}

For each of the NGP and SGP regions, 22 mock catalogues were generated
from the $\lcdm$ Hubble Volume simulation (Evrard \etal~2002) using
the techniques described in Cole \etal~(1998), and are designed to
have a similar clustering signal to the 2dFGRS. A summary of the
construction methods is presented here but for more details see
Norberg \etal~(2002a) and Baugh \etal~(in preparation).

These simulations used an initial dark matter power spectrum
appropriate to a flat $\lcdm$ model with $\om = 0.3$ and $\oml =
0.7$. The dark-matter evolution was followed up to the present day and
then a bias scheme (Model 2 of Cole \etal~1998, with a smoothing
length, $R_S = 2\hmpc$) was used to identify galaxies from the dark
matter haloes. The bias scheme used has two free parameters which are
adjusted to match the mean slope and amplitude of the correlation
function on scales greater than a few megaparsec. On scales smaller
than the smoothing length there is little control over the form of the
clustering, but in reality the methods employed work reasonably well
(see later Sections).

The resulting catalogues have a bias scheme which asymptotes to a
constant on large scales, giving $\beta = 0.47$, but is scale
dependent on small scales. Apparent magnitudes were assigned to the
galaxies consistent with their redshift, the assumed Schechter
luminosity function and the magnitude limit of the survey. The
Schechter function has essentially the same parameters as in the real
data (see Norberg \etal~2002a). Then the completeness mask and
variable apparent magnitude limits were applied to the mock catalogues
to reproduce catalogues similar to the real data.
 
In the analysis which follows, we make use of the real- and
redshift-space correlation functions from the full Hubble Volume
simulation. These correlation functions are determined from a Fourier
transform of the power spectrum of the full Hubble Volume cube using
the real- and redshift-space positions of the mass particles
respectively, along with the bias scheme outlined above. This allows
us to compare our mock catalogue results with that of the simulation
from which they are drawn to ensure we can reproduce the correct
parameters. It also allows us to compare and contrast the results from
the real Universe with a large numerical simulation.

\subsection{Error estimates}

We analyse each of the mock catalogues in the same way as the real
data, so that we have 22 mock measurements for every measurement that
we make on the real data. The standard deviation between the 22 mock
measurements gives a robust estimate of the uncertainty on the real
data. We use this approach to estimate the uncertainties for direct
measurements from the data, such as the individual points in the
correlation function, and for best-fit parameters such as $s_0$.

When fitting parameters we use this standard deviation as a weight
for each data-point and perform a minimum $\chi^2$ analysis to obtain
the best-fit parameter. The errors that we quote for any particular
parameter are the rms spread between the 22 best-fit parameters
obtained in the same way from the mock catalogues. This simple way of
estimating the uncertainties avoids the complications of dealing directly
with correlated errors in measured data points, while still providing
an unbiased estimate of the real uncertainties in the data, including
the effects of correlated errors.

Although this approach gives reliable estimates of the uncertainties,
the simple weighting scheme is not necessarily optimal in the presence
of correlated errors.  Nevertheless, for all statistics that we
consider, we find that the means of the mock estimates agree well with
the values input to the parent simulations. Also, we have applied the
technique described by Madgwick \etal\ (2003) to de-correlate the
errors for the projected correlation function, using the covariance
matrix estimated from the mock catalogues. We found a $0.1\sigma$
difference between the best-fit values using the two methods. So, we
are confident that our measurements and uncertainty estimates are
robust and unbiased.

\section{Estimates of the correlation function}

The two point correlation function, $\xi$, is measured by comparing
the actual galaxy distribution to a catalogue of randomly distributed
galaxies. These randomly distributed galaxies are subject to the same
redshift, magnitude and mask constraints as the real data and we
modulate the surface density of points in the random catalogue to
follow the completeness variations. We count the pairs in bins of
separation along the line-of-sight, $\pi$, and across the
line-of-sight, $\sigma$, to estimate $\xisp$. Spherically averaging
these pair counts provides the redshift-space correlation function
$\xis$. Finally, we estimate the projected function $\Xi(\sigma)$ by
integrating over all velocity separations along the line-of-sight and
invert it to obtain $\xir$.

\subsection{Constructing a random catalogue}

To reduce shot noise we compare the data with a random catalogue
containing ten times as many points as the real catalogue. This random
catalogue needs to have a smooth selection function matching the
$N(z)$ of the real data. We use the 2dFGRS luminosity function
(Norberg \etal~2002a) with $M^\ast_{\bj} - 5\log h = -19.66$ and $\alpha =
-1.21$ to generate the selection function, following the change in the
survey magnitude limit across the sky.  When analysing the mock
catalogues, we use the input luminosity function to generate the
selection function, and hence random catalogues.

As an alternative method, we also fitted an analytic form for the
selection function (Baugh \& Efstathiou 1993) to the data, and
generated random catalogues using that selection function.
We have calculated all of our statistics using both approaches, and
found that they gave essentially identical results for the data.  When
analysing the mock catalogues, we found that the luminosity function
method was more robust to the presence of large-scale features
in the $N(z)$ data. Thus, all of our quoted results are based on
random catalogues generated using the luminosity function.

\subsection{Fibre collisions}

The design of the 2dF instrument means that fibres cannot be placed
closer than approximately 30 arcsec (Lewis \etal~2002), and so both
members of a close pair of galaxies cannot be targeted in a single
fibre configuration. Fortunately, the arrangement of 2dFGRS tiles
means that not all close pairs are lost from the survey.  Neighbouring
tiles have significant areas of overlap, and so much of the sky is
targeted more than once. This allows us to target both galaxies in
some close pairs. Nevertheless, the survey misses a large fraction of
close pairs. It is important to assess the impact of this omission on
the measurement of galaxy clustering and to investigate schemes that
can compensate for the loss of close pairs.

\begin{figure}
\psfig{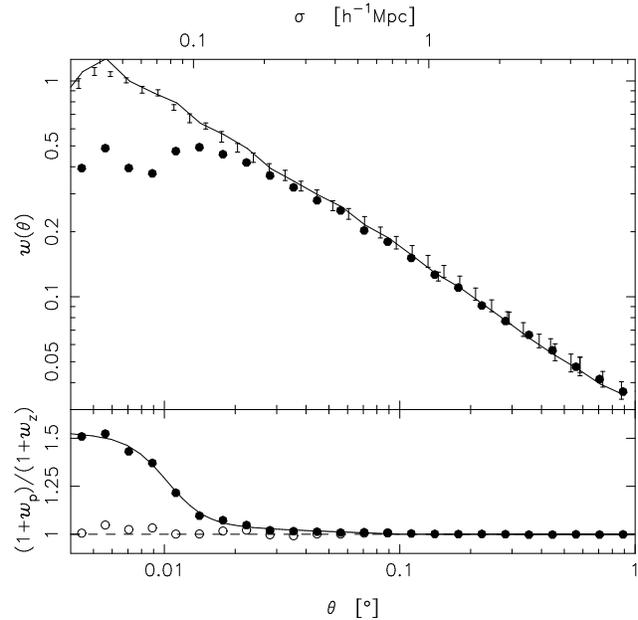}
\caption[]{Top panel: $w(\theta)$ for the mean of the NGP and SGP
redshift catalogues (solid points), the mean of the masked parent
catalogues (solid line), and the full APM result (error bars). Bottom
panel: The parent catalogue result divided by the redshift catalogue
results (uncorrected - solid points; collision corrected (see Section\
\ref{s:weight}) - open points). The solid line is the curve used to
correct the fibre collisions. The top axis converts $\theta$ into a
projected separation, $\sigma$, at the effective redshift of the
survey, $z_s = 0.15$.}
\label{f:ang}
\end{figure}

\begin{figure*}
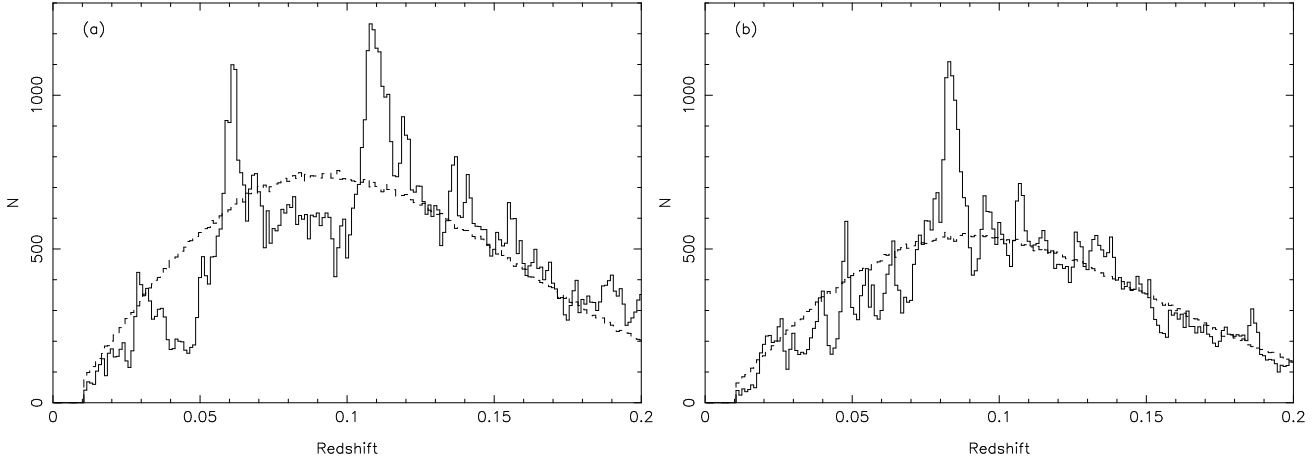

\parbox{\textwidth}{\hbox{
\psfig{figure=./nz_sgp.ps,angle=-90,width=8.6cm,clip=}
\psfig{figure=./nz_ngp.ps,angle=-90,width=8.6cm,clip=}}}
\caption[]{Redshift distributions, $N(z)$, for the 2dFGRS data (solid
lines) and the normalised random catalogues generated using the survey
luminosity function (dashed lines) for the (a) SGP and (b) NGP.}
\label{f:nzs}
\end{figure*}

To quantify the effect of these so-called `fibre collisions' we have
calculated the angular correlation function for galaxies in the 2dFGRS
parent catalogue, $w_p(\theta)$, and for galaxies with redshifts used
in our $\xi$ analysis, $w_z(\theta)$. We used the same mask to
determine the angular selection and apparent magnitude limit for each
sample as in Fig.\ \ref{f:masks}. Note that the mask is used only to
define the area of analysis, and the actual redshift completeness
values are not used in the calculation of $w$. In our $\xi$ analyses
we impose redshift limits $0.01 < z < 0.2$, which means that the mean
redshift of the redshift sample is lower than the parent sample. We
used Limber's equation (Limber 1954) to calculate the scale factors in
amplitude and angular scale needed to account for the different
redshift distributions. The solid line in Fig.\ \ref{f:ang} shows
$w_p$, and the filled circles show $w_z$ after applying the Limber
scale factors.  The error bars in Fig.\ \ref{f:ang} show $w(\theta)$
from the full APM survey (Maddox, Efstathiou \& Sutherland 1996), also
scaled to the magnitude limit of the 2dFGRS parent sample. On scales
$\theta \gtrsim 0.03^{\circ}$ all three measurements are
consistent. On smaller scales $w_z$ is clearly much lower than $w_p$,
showing that the fibre collision effect becomes significant and cannot
be neglected.

The ratio of galaxy pairs counted in the parent and redshift samples
is given by $(1+w_p)/(1+w_z)$, which is shown by the filled circles in
the lower panel of Fig.\ \ref{f:ang}. As discussed in the next section,
we use this ratio to correct the pair counts in the $\xi$ analysis.

\subsection{Weighting}
\label{s:weight}

Each galaxy and random galaxy is given a weighting factor depending on
its redshift and position on the sky. The redshift dependent part of
the weight is designed to minimize the variance on the estimated $\xi$
(Efstathiou 1988; Loveday \etal\ 1995), and is given by $1/(1 + 4 \pi
n(z_i) J_3(s))$, where $n(z)$ is the density distribution and $J_3(s)
= \int_0^s \xi(s')s'{^2} ds'$. We use $n(z)$ from the random catalogue
to ensure that the weights vary smoothly with redshift.  We find that
our results are insensitive to the precise form of $J_3$ but we
derived it using a power law $\xi$ with $s_0 = 13.0$ and $\gamma_s =
0.75$ and a maximum value of $J_3 = 400$.  This corresponds to the
best-fit power law over the range $0.1 <s < 3 \hmpc$ with a cutoff at
larger scales.

We also use the weighting scheme to correct for the galaxies that are
not observed due to the fibre collisions.  Each galaxy-galaxy pair is
weighted by the ratio $w_{f}=(1+w_p)/(1+w_z)$ at the relevant angular
separation according to the curve plotted in the bottom panel of Fig.\
\ref{f:ang}.
This corrects the observed pair count to what would have been counted
in the parent catalogue. The open points in Fig.\ \ref{f:ang}, which
have the collision correction applied, show that this method can
correctly recover the parent catalogue result and hence overcome the
fibre collision problem. Since the random catalogues do not have any
close-pair constraints, only the galaxy-galaxy pair count needs
correcting in this way.  We also tried an alternative approach to the
fibre-collision correction that we used previously in Norberg
\etal~(2001, 2002b) where the weight for each unobserved galaxy was
assigned equally to its ten nearest neighbours. This produced similar
results for $\theta>0.03^\circ$, but did not help on smaller
scales. All of our results are presented using the $w_{f}$ weighting
scheme. Hence each galaxy, $i$, is weighted by the factor,
\begin{equation}
w_i = \frac{1}{1 + 4 \pi n(z_i) J_3(s)}, 
\end{equation}
and each galaxy-galaxy pair $i$,$j$ is given a weight $w_f w_i w_j$,
whereas each galaxy-random and random-random pair is given a weight $w_i
w_j$.

\subsection{The two-point correlation function, $\xisp$} 

We use the $\xi$ estimator of Landy \& Szalay (1993),
\begin{equation}
\xisp  = \frac{DD - 2DR + RR}{RR}
\end{equation}
where $DD$ is the normalised sum of weights of galaxy-galaxy pairs with
particular $(\sigma, \pi)$ separation, $RR$ the normalised sum of
weights of random-random pairs with the same separation in the random
catalogue and $DR$ the normalised sum of weights of galaxy-random pairs
with the same separation. To normalise the pair counts we ensure that
the sum of weights of the random catalogue equal the sum of weights of
the real galaxy catalogue, as a function of scale. We find that other
estimators (\eg Hamilton 1993) give similar results.

The $N(z)$ distributions for the data and random catalogues (scaled so
that the area under the curve is the same as for the observed data)
are shown in Fig.\ \ref{f:nzs}.  It is clear that $N(z)$ for the
random catalogues are a reasonably smooth fit to $N(z)$ for the
data. Norberg \etal~(2002a) showed that large `spikes' in the $N(z)$
are common in the mock catalogues, and so similar features in the data
redshift distributions indicate normal structure.

\begin{figure*}
\psfig{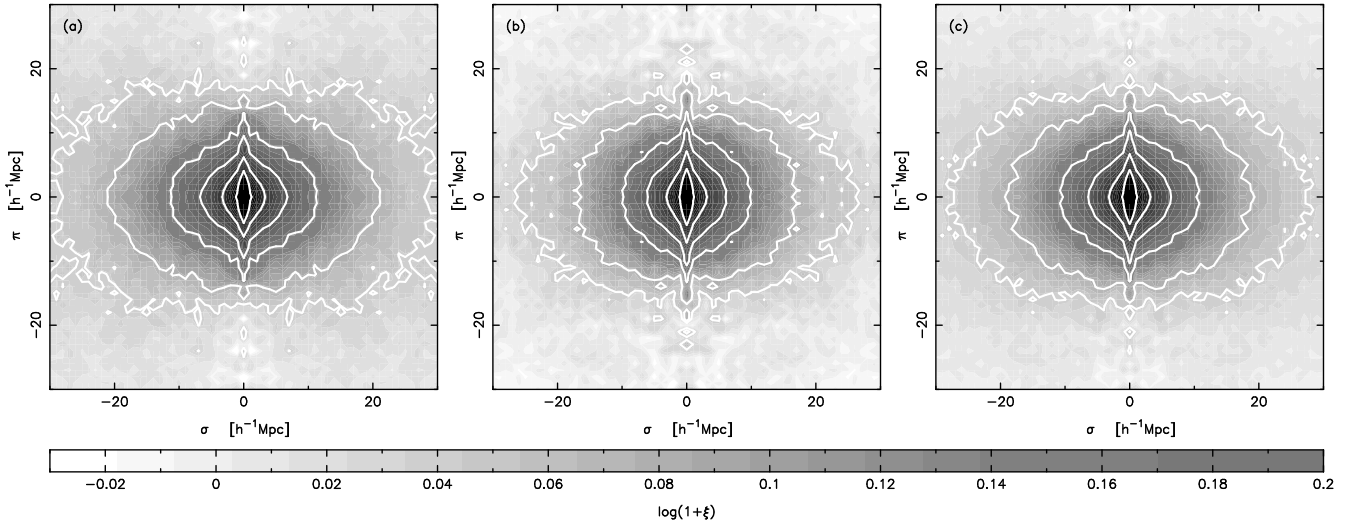}
\caption[]{Grey-scale plots of the 2dFGRS $\xisp$ (in $1\hmpc$ bins)
for (a) the SGP region, (b) the NGP region and (c) the combined
data. Contours are overlaid at $\xi=4.0, 2.0, 1.0, 0.5, 0.2$ and
$0.1$.}
\label{f:sigpi}
\end{figure*}

The resulting estimates of $\xisp$ calculated separately for the SGP
and NGP catalogues are shown in Fig.\ \ref{f:sigpi}, along with the
combined result.  The velocity distortions are clear at both small and
large scales, and the signal-to-noise ratio is in general very high
for $\sigma$ and $\pi$ values less than $20\hmpc$; it is $\approx 6$ in
each $1\hmpc$ bin at $s = 20\hmpc$. At very large separations $\xisp$
becomes very close to zero, showing no evidence for features that
could be attributed to systematic photometric errors.

We used an earlier version of the 2dFGRS catalogue to carry out a less
detailed analysis of $\xisp$ (Peacock \etal~2001). The current
redshift sample has about 1.4 times as many galaxies, though more
importantly it is more contiguous, and the revised photometry has
improved the uniformity of the sample. Nevertheless our new results
are very similar to our earlier analysis, demonstrating the robustness
of our results. The current larger sample allows us to trace $\xi$ out
to larger scales with smaller uncertainties.  Also, in our present
analysis we analyse mock catalogues to obtain error estimates which
are more precise than the previous error approximation (see Section\
\ref{s:john}).

\begin{figure}
\psfig{figure=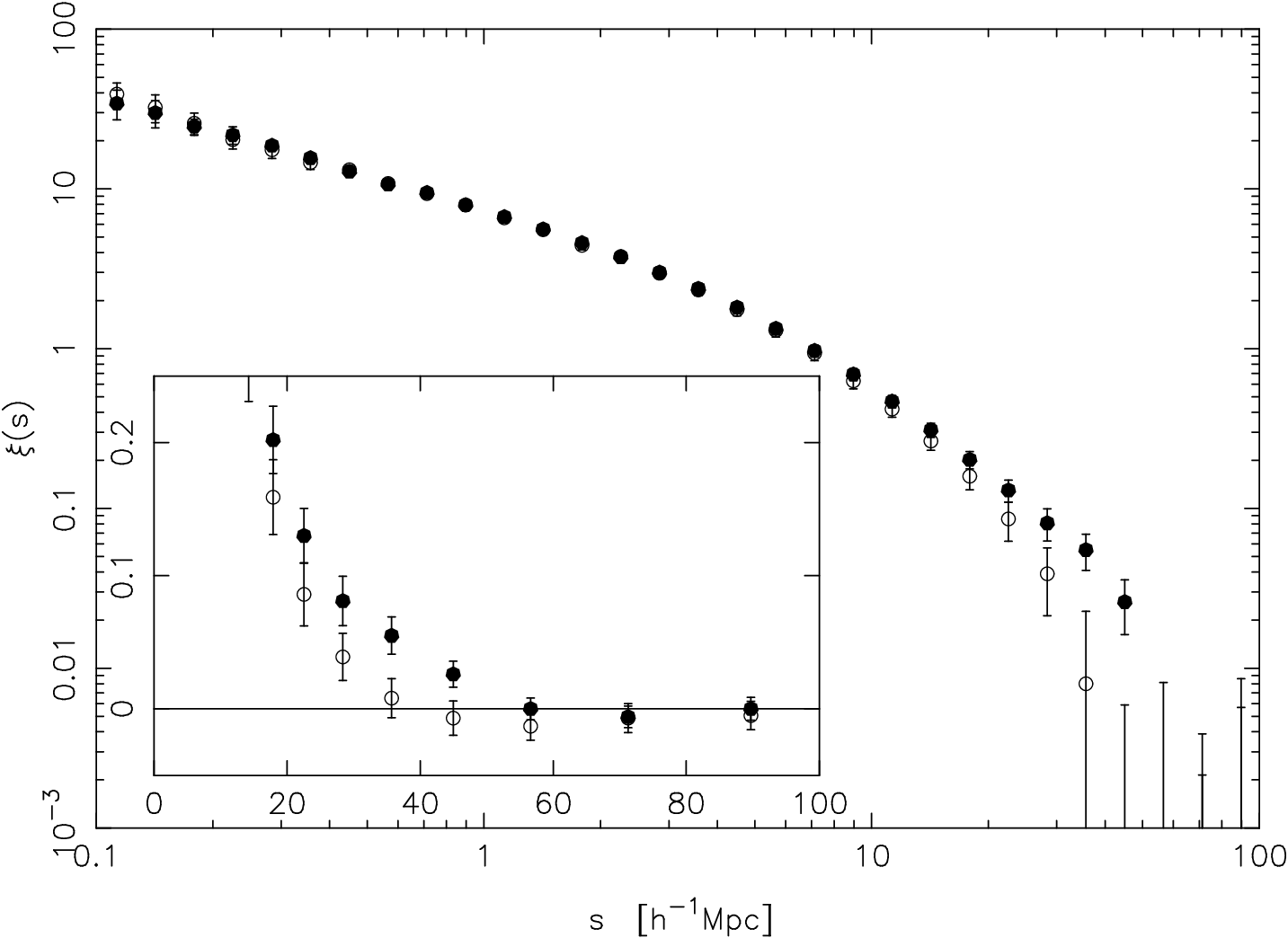,angle=-90,width=8.3cm,clip=}
\caption[]{The redshift-space correlation function for the NGP (open
points) and SGP (solid points) 2dFGRS data with error bars from the
rms of mock catalogue results. Inset plotted on a linear scale.}
\label{f:redxi}
\psfig{figure=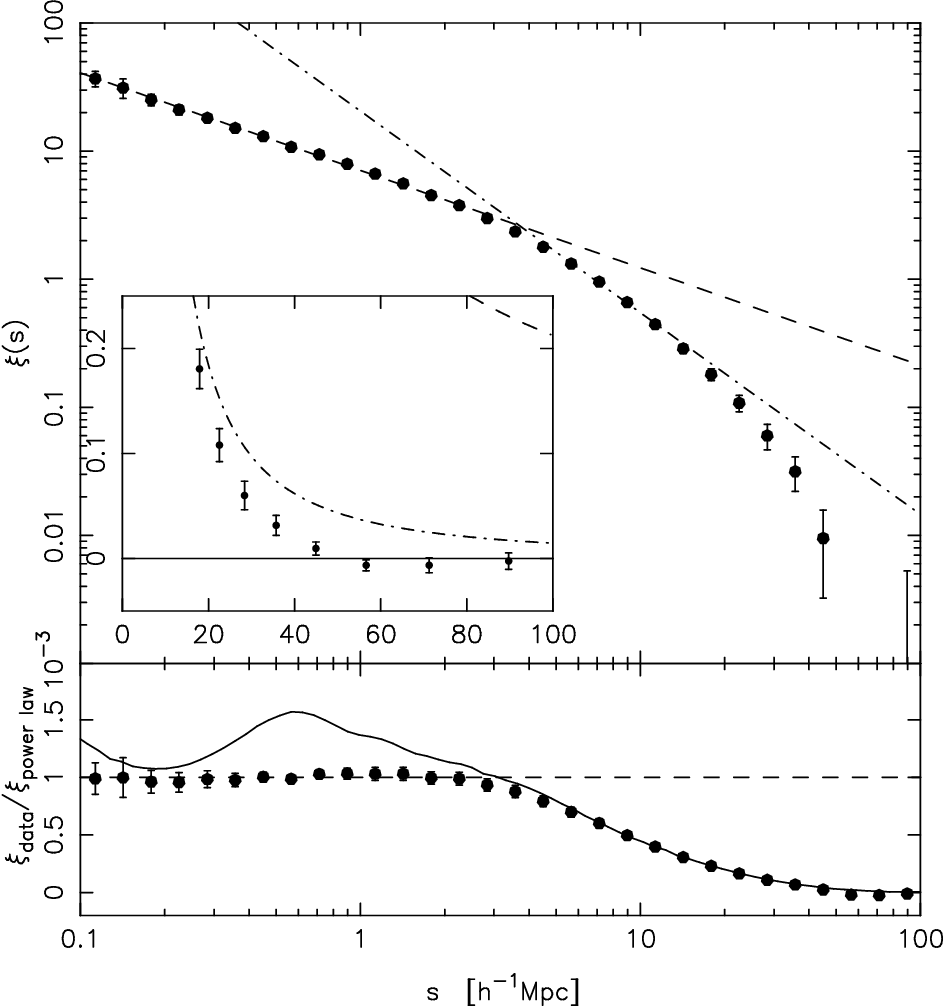,angle=0,width=8.3cm,clip=}
\caption[]{Top panel: The redshift-space correlation function for the
combined data (points) with error bars from the rms of the mock
catalogue results. The dashed line is a small scale power law fit,
($s_0 = 13\hmpc$, $\gamma_s = 0.75$) and the dot-dashed line is the
best-fit to points around $s_0$, ($s_0 = 6.82\hmpc$, $\gamma_s =
1.57$). Inset is on a linear scale. Bottom panel: As above, divided by
the small scale power law. The solid line shows the result from the
Hubble Volume simulation.}
\label{f:redxi2}
\end{figure}

\subsection{The redshift-space correlation function, $\xis$}
\label{s:red}

Averaging $\xisp$ at constant $s$ gives the redshift-space correlation
function, and our results for the NGP and SGP are plotted in Fig.\
\ref{f:redxi} on both log and linear scales. The NGP and SGP
measurements differ by about $2\sigma$ between $20$ and $50\hmpc$, and
we find one mock whose NGP and SGP measurements disagree by this much,
and so it is probably not significant. We tried shifting $M^{\ast}$ by
$0.1\;$mag to better fit the $N(z)$ at $z > 0.15$ in the SGP, and this
moved the data points by $\sim 0.2\sigma$ for $20 < s < 50\hmpc$.

The redshift-space correlation function for the combined data is
plotted in Fig.\ \ref{f:redxi2} in the top panel. It is clear that the
measured $\xis$ is not at all well represented by a universal power
law on all scales, but we do make an estimate of the true value of the
redshift-space correlation length, $s_0$, by fitting a localised
power-law of the form,
\begin{equation}
\label{e:s0}
\xis = \left (\frac{s}{s_0} \right )^{-\gamma_s}
\end{equation}
using a least-squares fit to $\log(\xi)$ as a function of $\log(s)$,
using two points either side of $\xis = 1$. This also gives a value
for the local redshift-space slope, $\gamma_s$. The best-fit
parameters for the separate poles and combined estimates are listed in
Table\ \ref{t:bf}. In the inset of Fig.\ \ref{f:redxi2} we can see, at
a low amplitude, that $\xi(s)$ goes negative between $50 \lesssim s
\lesssim 90\hmpc$.

In the bottom panel of Fig.\ \ref{f:redxi2} we examine the shape of
$\xis$ more carefully. The points are the data divided by a small
scale power law fitted on scales $0.1 < s < 3\hmpc$ (dashed line).
The data are remarkably close to the power-law fit for this limited
range of scales, and follow a smooth break towards zero for $3 < s <
60\hmpc$. The measurements from the Hubble Volume simulation are shown
by the solid line, and it matches the data extremely well on scales $s
> 4\hmpc$. On smaller scales, where the algorithm for placing galaxies
in the simulation has little control over the clustering amplitude (as
discussed in Section\ \ref{s:mocks}), there are discrepancies of order
$50\%$.

The mean $\xis$ determined from the mock catalogues agrees well with
the true redshift-space correlation function from the full Hubble
Volume. This provides a good check that our weighting scheme and
random catalogues have not introduced any biases in the analysis.

\subsection{Redshift-space comparisons}

Redshift-space correlation functions have been measured from many
redshift surveys, but direct comparisons between different surveys
are not straightforward because galaxy clustering depends on the
spectral type and luminosity of galaxies (\eg Guzzo \etal~2000;
Norberg \etal~2002b; Madgwick \etal~2003). Direct comparisons can be
made only between surveys that are based on similar galaxy selection
criteria. The 2dFGRS is selected using pseudo-total magnitudes in the
$\bj$ band, and the three most similar surveys are the Stromlo-APM
survey (SAPM, Loveday \etal~1992), the Durham UKST survey (Ratcliffe
\etal~1998) and the ESO Slice Project (ESP, Guzzo \etal~2000).  The
Las Campanas Redshift Survey (LCRS, Lin \etal~1996, Jing \etal~1998)
and Sloan Digital Sky Survey (SDSS, Zehavi \etal~2002) are selected in
the $R$ band, but have a very large number of galaxies, and so are
also interesting for comparisons.

\begin{figure}
\psfig{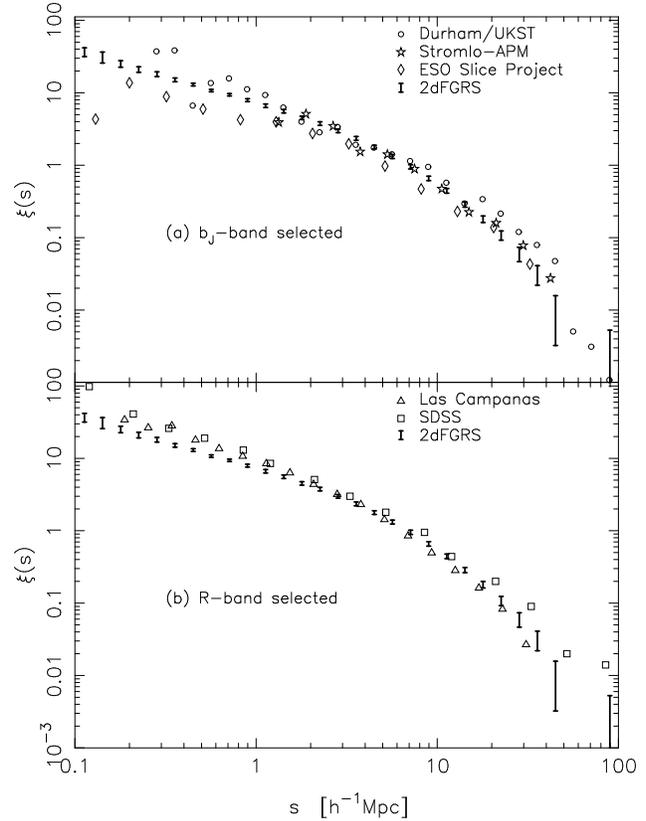}
\caption[]{Comparison of 2dFGRS $\xis$ with (a) other $\bj$ band
selected surveys as indicated and (b) $R$ band selected surveys as
indicated. These results are discussed in the text.}
\label{f:xis_comp}
\end{figure}

The non-power-law shape of $\xis$ makes it difficult to compare
different measurements of $s_0$ and $\gamma_s$, because the values
depend sensitively on the range of $s$ used in the fitting procedure.
In Fig.\ \ref{f:xis_comp}(a) we compare the $\xis$ measurements
directly for the 2dFGRS, SAPM, Durham UKST and ESP surveys.  Our
estimate of $\xis$ is close to the mean of previous measurements, but
the uncertainties are much smaller. Although we quote uncertainties
that are similar in size to previous measurements, we have used the
scatter between mock catalogues to estimate them, rather than the
Poisson or boot-strap estimates that have been used before and which
seriously underestimate the true uncertainties.

Fig.\ \ref{f:xis_comp}(b) shows the 2dFGRS measurements together with
the LCRS and SDSS measurements. On scales $s \gtrsim 4 \hmpc$ there
appears to be no significant differences between the surveys, but for
$s \lesssim 2 \hmpc$ the LCRS and SDSS have a higher amplitude than
the 2dFGRS. This difference is likely to be caused by the different
galaxy selection for the surveys, though the SDSS results shown are
for the Early Data Release (EDR) and have larger errors than the
2dFGRS points. The 2dFGRS is selected using $\bj$, whereas the SDSS
and LCRS are selected in red bands. Since the red (early type)
galaxies are more strongly clustered than blue (late type) galaxies
(\eg Zehavi \etal~2002; and via spectral type, Norberg \etal~2002b),
we should expect that $\xi$ will be higher for red selected surveys
than a blue selected survey. This issue is examined further in
Madgwick \etal~(2003).

\subsection{The projected correlation function, $\Xi(\sigma)$}
\label{s:pro}

The redshift-space correlation function differs significantly from the
real-space correlation function because of redshift-space distortions
(see Section 4). We can estimate the real-space correlation length,
$r_0$, by first calculating the projected correlation function,
$\Xi(\sigma)$. This is related to $\xisp$ via the equation,
\begin{equation}
\Xi(\sigma) = 2\int^{\infty}_{0}\xisp~d\pi
\label{e:Xi}
\end{equation}
though in practice we set the upper limit in this integral to
$\pi_{\rm{max}} = 70\hmpc$. The result is insensitive to this choice
for $\pi_{\rm{max}} > 60\hmpc$ for our data. Since redshift space
distortions move galaxy pairs only in the $\pi$ direction, and the
integral represents a sum of pairs over all $\pi$ values,
$\Xi(\sigma)$ is independent of redshift-space distortions.  It is
simple to show that $\Xi(\sigma)$ is directly related to the
real-space correlation function (Davis \& Peebles 1983),
\begin{equation}
\frac{\Xi(\sigma)}{\sigma} =
\frac{2}{\sigma}\int^{\infty}_{\sigma}\frac{r \xir dr}{(r^2 -
\sigma^2)^{\frac{1}{2}}}.
\label{e:proj}
\end{equation}
If the real-space correlation function is a power law this can be
integrated analytically. We write $\xir=(r/r^P_0)^{-\gamma^P_r}$,
where the $P$ superscripts refer to the `Projected' values, rather
than the `Inverted' values which are calculated in Section\ \ref{s:inv}
and denoted by $I$. With this notation we obtain,
\begin{equation}
\frac{\Xi(\sigma)}{\sigma} = \left(\frac{r_0^P}{\sigma}\right)^{\gamma_r^P}\frac{\Gamma(\frac{1}{2})\Gamma(\frac{\gamma^P_r - 1}{2})}{\Gamma(\frac{\gamma_r^P}{2})} = \left(\frac{r_0^P}{\sigma}\right)^{\gamma^P_r}A(\gamma^P_r).
\label{e:Xi2}
\end{equation}
The parameters $\gamma_r^P$ and $r_0^P$ can then be estimated from the
measured $\Xi(\sigma)$, giving an estimate of the real-space
clustering independent of any peculiar motions.

\begin{figure}
\psfig{figure=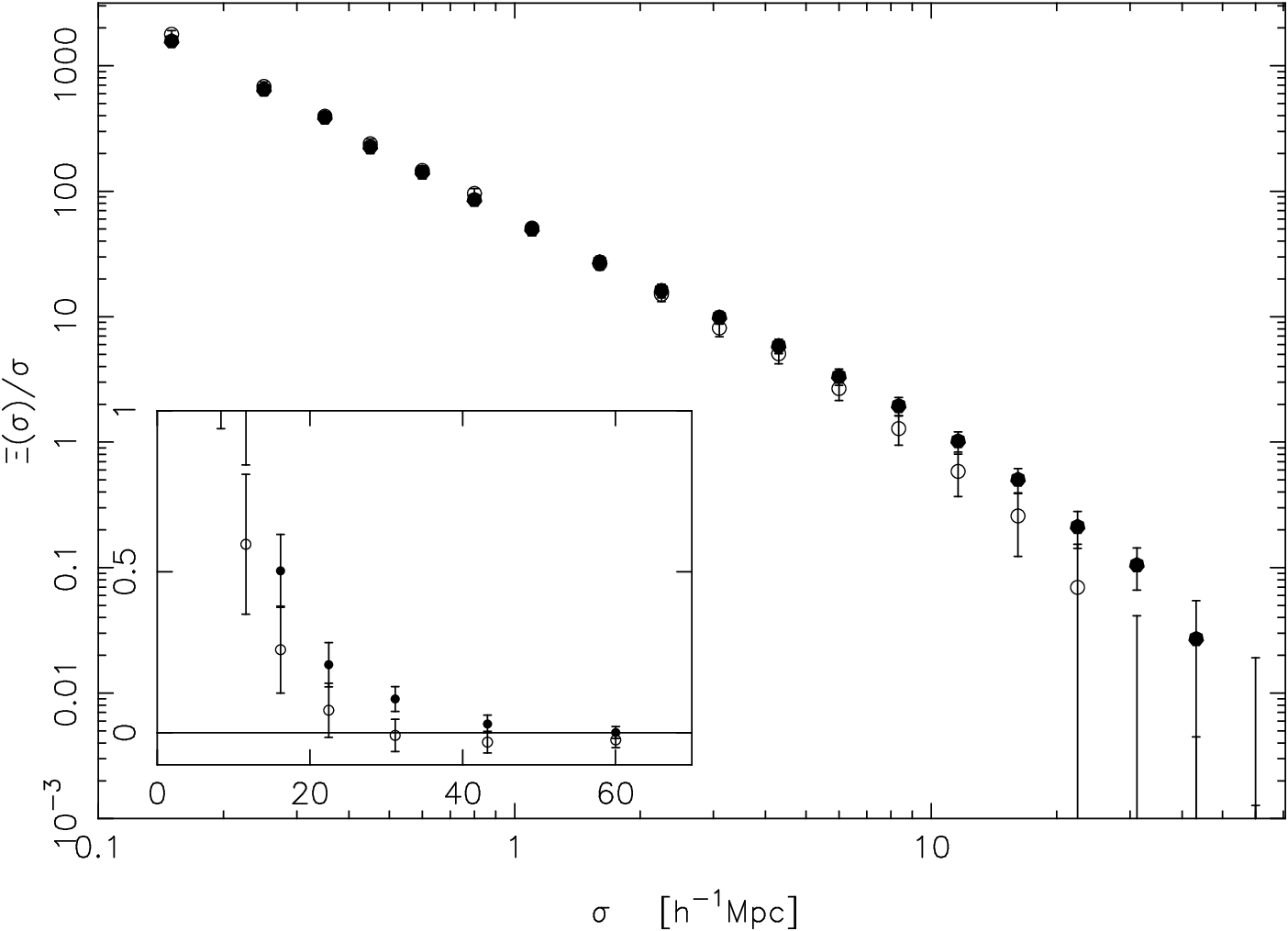,angle=-90,width=8.3cm,clip=}
\caption[]{The projected correlation functions for the NGP (open
points) and SGP (solid points) 2dFGRS data with error bars from the
rms spread between mock catalogue results. Inset plotted on a linear
scale.}
\label{f:Xi}
\psfig{figure=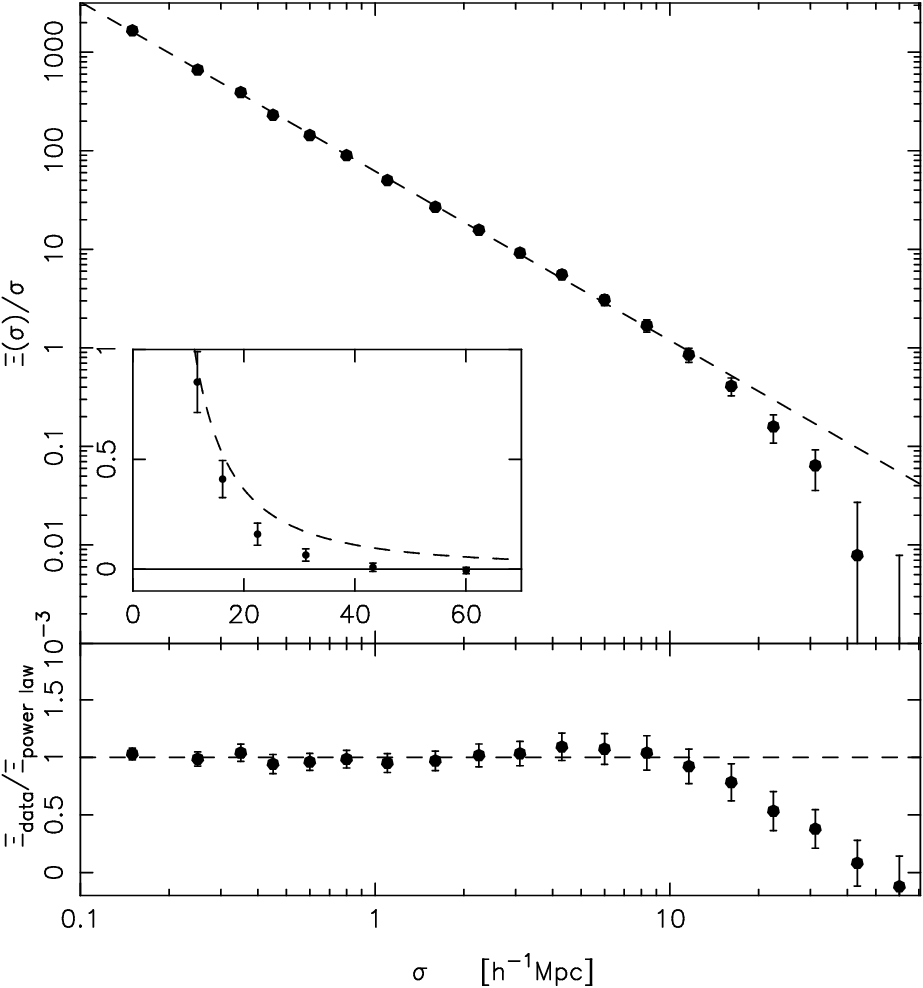,angle=0,width=8.3cm,clip=}
\caption[]{Top panel: The projected correlation function of the
combined data with error bars from the rms spread between mock
catalogue results. The dashed line is the best-fit power-law for $0.1
< \sigma < 12\hmpc$ ($r_0 = 4.98$, $\gamma_r = 1.72$, $A =
3.97$). Inset is plotted on a linear scale. Bottom panel: The combined
data divided by the power-law fit.}
\label{f:Xi2}
\end{figure}

The projected correlation functions for the NGP and SGP are shown in
Fig.\ \ref{f:Xi} and the combined data result is shown in Fig.\
\ref{f:Xi2}. The best-fit values of $\gamma_r^P$ and $r_0^P$ for $0.1
< \sigma < 12\hmpc$ are shown in Table\ \ref{t:bf}. Over this range
$\pcf$ is an accurate power law, but it steepens for $\sigma >
12\hmpc$.  This deviation from power-law behaviour limits the scales
that can be probed using this approach.

\subsection{The real-space correlation function, $\xir$}
\label{s:inv}

It is possible to estimate $\xir$ by directly inverting $\Xi(\sigma)$
without making the assumption that it is a power law (Saunders,
Rowan-Robinson \& Lawrence 1992, hereafter S92). They recast Eqn.\
\ref{e:proj} into the form,
\begin{equation}
\xir = -\frac{1}{\pi}\int^{\infty}_r \frac{(d\Xi(\sigma)/d\sigma)}{(\sigma^2 - r^2)^{\frac{1}{2}}}d\sigma.
\end{equation}
Assuming a step function for $\Xi(\sigma)=\Xi_i$ in bins centered on
$\sigma_i$, and interpolating between values,
\begin{equation}
\xi(\sigma_i) = -\frac{1}{\pi}\sum_{j\geq i}\frac{\Xi_{j+1}-\Xi_j}{\sigma_{j+1}-\sigma_j}\ln\left(\frac{\sigma_{j+1}+\sqrt{\sigma^2_{j+1} - \sigma^2_i}}{\sigma_j + \sqrt{\sigma^2_j - \sigma^2_i}}\right)
\end{equation}
for $r = \sigma_i$. S92 suggest that their method is only good for
scales $r \lesssim 30\hmpc$ in the QDOT survey because $r$ becomes
comparable to the maximum scale out to which they can estimate
$\Xi$. We can test the reliability of our inversion of the 2dFGRS data
using the mock catalogues.

\begin{figure}
\psfig{figure=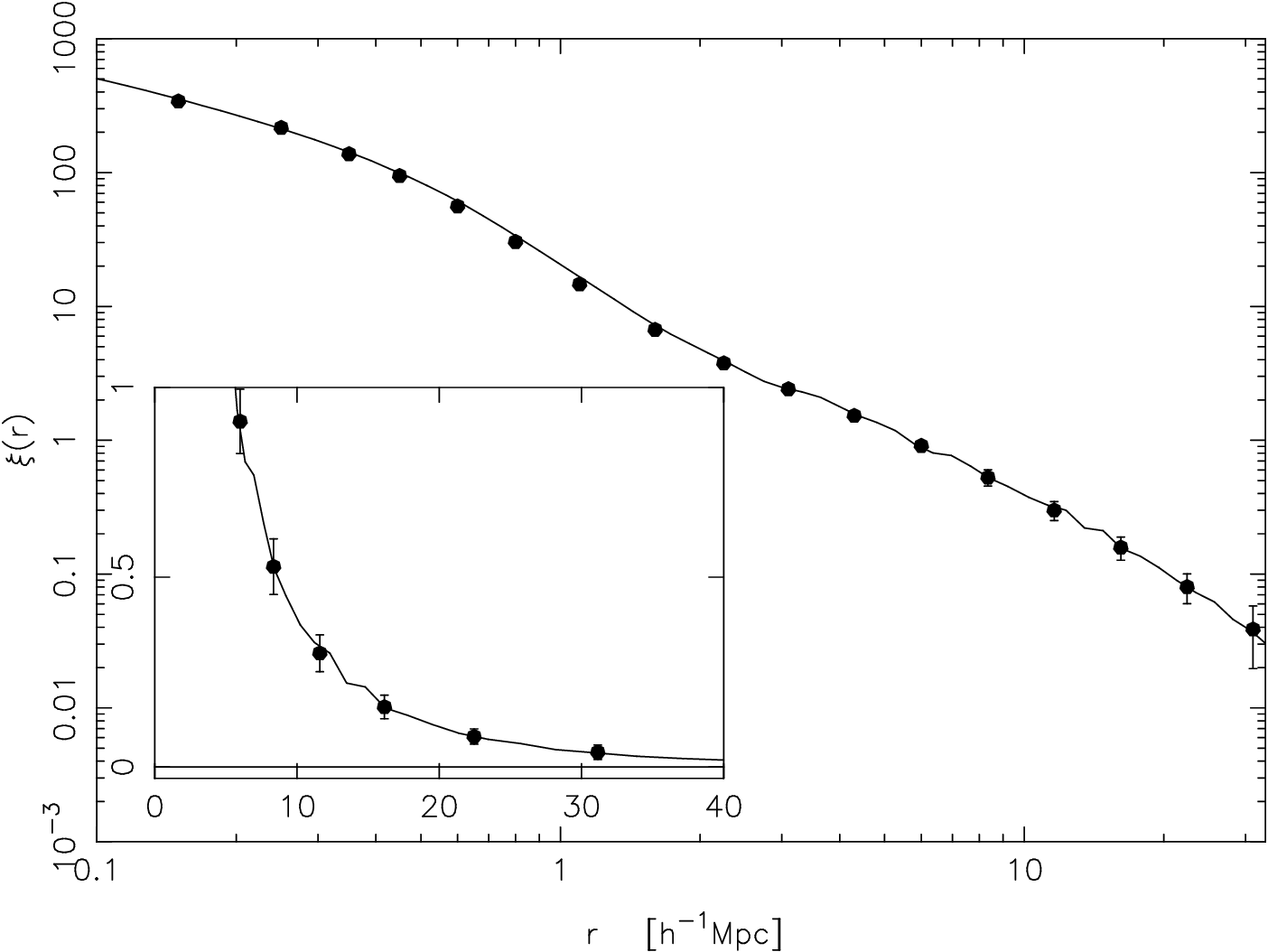,angle=-90,width=8.3cm,clip=}
\caption[]{The mean real-space correlation function determined from
the 22 mock catalogues using the method of S92. The solid line is the
true $\xir$ from the Hubble Volume and the agreement is
excellent. Note the changed scale from previous plots.}
\label{f:xir_mock}
\psfig{figure=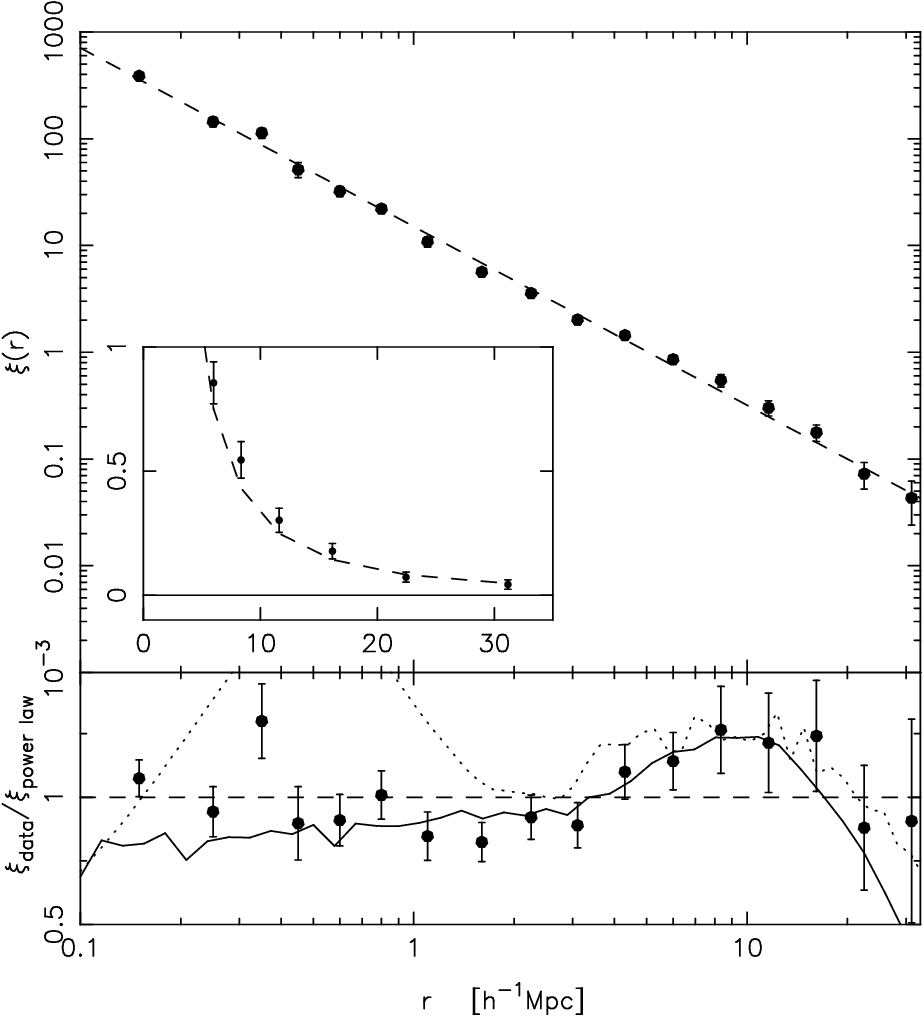,angle=0,width=8.3cm,clip=}
\caption[]{Top panel: The real-space correlation function of the
combined 2dFGRS using the method of S92, with error bars from the rms
spread between mock catalogues. The dashed line is the best-fit power
law ($r_0 = 5.05$, $\gamma_r = 1.67$). Inset is plotted on a linear
scale. Bottom panel: The data divided by the power law fit. The solid
line is the de-projected APM result (Padilla \& Baugh 2003) as
discussed in the text. The dotted line is the result from the Hubble
Volume.}
\label{f:xir}
\end{figure}

In Fig.\ \ref{f:xir_mock} we show the mean $\xir$ as determined from
the mock catalogues using the method of S92. We compare this to the
real-space correlation function determined directly from the Hubble
Volume simulation, from which the mock catalogues are drawn. The
agreement is excellent and shows that the method works and that we can
recover the real-space correlation function out to $30\hmpc$. Like S92
we find that beyond this scale the method begins to fail and the true
$\xir$ is not recovered.

We have applied this technique to the combined 2dFGRS data and obtain
the real-space correlation function shown in Fig.\ \ref{f:xir}. The
data are plotted out to only $30\hmpc$ due to the limitations in the
method described above. On small scales $\xir$ is well represented by
a power law, and a best-fit over the range $0.1 < r < 12\hmpc$ gives
the results for $r_0^I$ and $\gamma_r^I$ shown in Table\ \ref{t:bf}.

The points in the bottom panel of Fig.\ \ref{f:xir} show the 2dFGRS
data divided by the best-fit power law. It can be seen that at scales
$0.1 < r < 20\hmpc$ the data $\xir$ is close to the best-fit power-law
but does show hints of non power-law behaviour (see also discussion
below). 

\begin{table}
\caption{Best-fit parameters to $\xi$. For $s_0$ and $\gamma_s$ the
fit to $\xis$ uses only points around $s=s_0$. For $r_0^P$,
$\gamma_r^P$ and A($\gamma_r^P$) the fit to $\pcf$ uses all points
with $0.1 < \sigma < 12\hmpc$. For $r_0^I$ and $\gamma_r^I$ the fit to
the inverted $\xir$ uses all points with $0.1 < r < 12\hmpc$. In each
case the errors quoted are the rms spread in the results obtained from
the same analysis with the mock catalogues.}
\label{t:bf}
\begin{tabular}{llll}
\hline
Parameter       & SGP               & NGP               & Combined\\
\hline		
$s_0~(\hmpcb)$  & $6.92 \pm 0.36$ & $6.72 \pm 0.41$ & $\as0$\\
$\gamma_s$      & $1.51 \pm 0.08$ & $1.64 \pm 0.08$ & $\agams$\\
\hline		
$r_0^P~(\hmpcb)$& $5.05 \pm 0.32$ & $4.79 \pm 0.31$ & $4.95 \pm 0.25$\\
$\gamma_r^P$    & $1.68 \pm 0.06$ & $1.77 \pm 0.07$ & $1.72 \pm 0.04$\\
A($\gamma_r^P$) & $4.17 \pm 0.23$ & $3.77 \pm 0.28$ & $3.99 \pm 0.16$\\
\hline		
$r_0^I~(\hmpcb)$& $5.09 \pm 0.35$ & $5.08 \pm 0.28$ & $\ar0$\\
$\gamma_r^I$    & $1.65 \pm 0.03$ & $1.70 \pm 0.04$ & $\agamr$\\
\hline
\end{tabular}
\end{table}

\subsection{Real-space comparisons}
\label{s:xir_comp}

In the inverted $\xir$ (and possibly $\Xi[\sigma]$) there is a weak
excess of clustering over the power-law for $5 < r < 20 \hmpc$. This
has been previously called a `shoulder' in $\xi$ (see \eg Ratcliffe
\etal~1998). Though the amplitude of the feature in our data is rather
low, it has been consistently seen in different surveys, and probably
is a real feature. After submission of this work, Zehavi \etal\ (2003)
also saw this effect in the SDSS projected correlation function and
explained the inflection point as the transition scale between a
regime dominated by galaxy pairs in the same halo and a regime
dominated by pairs in separate haloes. Magliochetti \& Porciani (2003)
have found the same effect when examining correlation functions of
different types of 2dFGRS galaxy.

The dotted line in the bottom panel of Fig.\ \ref{f:xir} shows the
Hubble Volume simulation which agrees well with the data for $r >
1\hmpc$. On smaller scales the Hubble Volume $\xi$ shows significant
deviations from a power-law. On these scales, the galaxy clustering
amplitude in the simulation is incorrectly modelled since the
assignment of galaxies to particles is based on the mass distribution
smoothed on a scale of $2\hmpc$ (as discussed in Section\
\ref{s:mocks}). The solid line in the bottom panel of Fig.\
\ref{f:xir} is the de-projected APM result (Padilla \& Baugh 2003),
scaled down by a factor $(1+z_s)^{\alpha}$, with $\alpha = 1.7$,
suitable for evolution in a $\lcdm$ cosmology. There is good agreement
between the 2dFGRS and APM results which are obtained using quite
different methods.

We have estimated $r_0$ and $\gamma_r$ by fitting to the projected
correlation function $\pcf$, and also by inverting $\pcf$ and then
fitting to $\xir$. The best-fit values from the two methods are shown
in Table\ \ref{t:bf}, and it is clear they lead to very similar
estimates of $r_0$ and $\gamma_r$. This confirms that the power-law
assumption in Section\ \ref{s:pro} is a good approximation over the
scales we consider.

\begin{table}
\caption{Measurements of $\xir$ from 2dFGRS and other surveys, with
the quoted uncertainties as published. The various authors have used
very different ways to estimate errors though none have included the
effects of cosmic variance. Since we have used the scatter between
mock catalogues to take account of this, our estimates are actually
dominated by cosmic variance. These results are measured at different
effective luminosities, redshifts and for different galaxy types.}
\label{t:bf_comp}
\begin{tabular}{lll}
\hline
Survey      & $r_0~(\hmpcb)$  & $\gamma_r$ \\
\hline
2dFGRS ($P$)& $4.95 \pm 0.25$ & $1.72 \pm 0.04$ \\
2dFGRS ($I$)& $\ar0$          & $\agamr$  \\
SAPM        & $5.1\pm0.3$     & $1.71\pm0.05$ \\
ESP         & $4.15\pm 0.2$   & $1.67^{+0.07}_{-0.09}$ \\
Durham UKST & $5.1\pm0.3$     & $1.6 \pm 0.1$\\
LCRS        & $5.06\pm0.12$   & $1.86\pm0.03$\\
SDSS        & $6.14\pm0.18$   & $1.75\pm0.03$\\
\hline
\end{tabular}
\end{table}

Table\ \ref{t:bf_comp} lists $r_0$ and $\gamma_r$ for the 2dFGRS and
other surveys estimated using power-law fits to the projected
correlation function $\Xi(\sigma)$. As mentioned in Section\
\ref{s:red} the SAPM, Durham UKST and the ESP are $\bj$ selected
surveys, and so should be directly comparable to the 2dFGRS. The
values of $r_0$ and $\gamma_r$ for these surveys all agree to within
one standard deviation, except $r_0$ for the ESP, which appears to be
significantly lower. It is likely that the quoted uncertainties for
the ESP and Durham UKST parameters are underestimated since they did
not include the effect of cosmic variance. Since they each sample
relatively small volumes, this will be a large effect.  The sparse
sampling strategy used in the SAPM means that it has a large effective
volume, and so the cosmic variance is small.

As in Section\ \ref{s:red}, the red-selected surveys, LCRS and SDSS,
are significantly different from the other surveys. The discrepancies
are most likely due to the fact that the amplitude of galaxy
clustering depends on galaxy type, and that red-selected surveys have
a different mix of galaxy types. We can make a very rough
approximation of the expected change in $\xi$ by considering how the
mean colour difference of early and late populations changes the
relative fraction of the two populations when a magnitude limited
sample is selected in different pass bands. Zehavi \etal\ (2002) split
their $r-$selected SDSS sample into 19603 early-type galaxies and 9532
late-type galaxies. The mean $(g-r)$ colours are 0.5 and 0.9
respectively. The 2dFGRS is selected using $\bj$ which is close to
$g$, and so compared to the $r$ selection, the median depth for blue
galaxies will be larger that for red galaxies. The number of early and
late types will roughly scale in proportion to the volumes sampled,
and so the ratio of early-to-late galaxies in the 2dFGRS will be
roughly $\sim (19603/9532) \times 10^{0.6(0.5-0.9)} = 1.18$.  Note
that this colour split leads to a very different ratio of
early-to-late galaxies compared to the $\eta$ split used by Madgwick
\etal\ (2003). Assuming the early and late correlation functions trace
the same underlying field, the combined correlation function will be
\begin{equation}
\xi_{\rm tot} = \left({n_{\rm early}b_{\rm early}+n_{\rm late}b_{\rm
late} \over n_{\rm early}+n_{\rm late}}\right)^2 \xi_{\rm mass}.
\end{equation}
From the power law fits of Zehavi \etal, the ratio of bias values at 1
Mpc is $b_{\rm early}/b_{\rm late} = 4.95$.  Inserting the different
ratios $n_{\rm early}/n_{\rm late}$ appropriate to the red and blue
selected samples we find that the expected ratio of $\xi$ for a red
selected sample compared to a blue selected sample is roughly
1.36. Scaling the 2dFGRS values of $r_0 = 5.05$ and $\gamma_r = 1.67$
leads to a SDSS value of $r_0 = 5.95$ for $\gamma_r = 1.75$, within
$1\sigma$ of the actual SDSS value. This simple argument indicates
that the observed difference in $\xi$ between the red and blue
selected surveys is consistent with the different population mixes
expected in the surveys. The extra surface brightness selection
applied to the LCRS may also introduce significant biases.

Each survey is also likely to have a different effective luminosity
and, as has been shown by Norberg \etal~(2001), this will cause
clustering measurements to differ. The relation for 2dFGRS galaxies
found by Norberg \etal~(2001) was,
\begin{equation}
\left(\frac{r_0}{r_0^{\ast}}\right)^{\nts\nts\Large\frac{\gamma_r}{2}} = 0.85 + 0.15\left(\frac{L}{L^{\ast}}\right),
\label{e:r0star}
\end{equation}
which gives, for $L = 1.4L^{\ast}$ (see Section\ \ref{s:2dfdata}),
$r_0^{\ast} = 4.71 \pm 0.24$, which will allow direct comparisons with
other surveys.

\section{Redshift-space distortions}
\label{s:dis}

When analysing redshift surveys it must be remembered that the
distance to each galaxy is estimated from its redshift and is not the
true distance. Each galaxy has, superimposed on its Hubble motion, a
peculiar velocity due to the gravitational potential in its local
environment. These peculiar velocities can be in any direction and,
since this effect distorts the correlation function, it can be used to
measure two important parameters.

The peculiar velocities are caused by two effects. On small scales,
random motions of the galaxies within groups cause a radial smearing
known as the `Finger of God'. On large scales gravitational
instability leads to coherent infall into overdense regions and
outflow from underdense regions. We analyse the observed
redshift-space distortions by modeling $\xisp$. We start with a model
of the real-space correlation function, $\xir$, and include the
effects of large-scale coherent infall, which is parameterized by
$\beta\approx\om^{0.6}/b$, where $b$ is the linear bias parameter. We
then convolve this with the form of the random pairwise motions.

\subsection{Constructing the model}

Kaiser (1987) pointed out that, in the linear regime, the coherent
infall velocities take a simple form in Fourier space. Hamilton (1992)
translated these results into real space,
\begin{equation}
\xi'(\sigma, \pi) = \xi_0(s)P_0(\ct) + \xi_2(s)P_2(\ct) + \xi_4(s)P_4(\ct)
\end{equation}
where P$_{\ell}(\mu)$ are Legendre polynomials, $\ct = \cos(\theta)$
and $\theta$ is the angle between $r$ and $\pi$. The relations between
$\xi_\ell$, $\xir$ and $\beta$ for a simple power-law
$\xir=(r/r_0)^{-\gamma_r}$ are (Hamilton 1992),
\begin{equation}
\label{e:xi0}
\xi_0(s) = \left(1 + \frac{2\beta}{3} + \frac{\beta^2}{5}\right)\xir
\end{equation}
\begin{equation}
\xi_2(s) = \left(\frac{4\beta}{3} + \frac{4\beta^2}{7}\right)\left(\frac{\gamma_r}{\gamma_r - 3}\right)\xir
\end{equation}
\begin{equation}
\xi_4(s) = \frac{8\beta^2}{35}\left(\frac{\gamma_r(2+\gamma_r)}{(3-\gamma_r)(5-\gamma_r)}\right)\xir.
\label{e:xi4}
\end{equation}
The Appendix has more details of this derivation and gives the
equations for the case of non-power law forms of $\xi$.

We use these relations to create a model $\xi'(\sigma, \pi)$
which we then convolve with the distribution function of random
pairwise motions, $f(v)$, to give the final model $\xisp$ (Peebles
1980):
\begin{equation}
\xisp = \int^{\infty}_{-\infty}\xi'(\sigma, \pi - v/H_0)f(v)dv
\end{equation} 
and we choose to represent the random motions by an exponential form,
\begin{equation}
f(v) = \frac{1}{a\sqrt{2}}\exp\left(-\frac{\sqrt{2}|v|}{a}\right)
\label{e:fv}
\end{equation} 
where $a$ is the pairwise peculiar velocity dispersion (often known as
$\sigma_{12}$). An exponential form for the random motions has been
found to fit the observed data better than other functional forms (\eg
Ratcliffe \etal~1998; Landy 2002; see also Section\ \ref{s:expvel}).

\subsection{Model assumptions}

In this model we make several assumptions. Firstly, we assume a
power-law for the correlation function.  The power-law approximation
is a good fit on scales $< 20\hmpc$ but is not so good at larger
scales. This limits the scales which we can probe using this method.
In Section\ \ref{s:gridfit}, we consider non-power-law models for
$\xir$, and recalculate Eqns.\ \ref{e:xi0} to \ref{e:xi4} using
numerical integrals (see Appendix), allowing us to reliably use scales
$> 20\hmpc$. Secondly, we assume that the linear theory model
described above holds on scales $\lesssim 8\hmpc$, which is almost
certainly not true. We also consider this in Section 7. Finally, we
assume an exponential distribution of peculiar velocities with a
constant velocity dispersion, $a$, (Eqn.\ \ref{e:fv}) and this is
discussed and justified in Section\ \ref{s:expvel} and Section\
\ref{s:pecvels}.

\subsection{Model plots}

\begin{figure}
\psfig{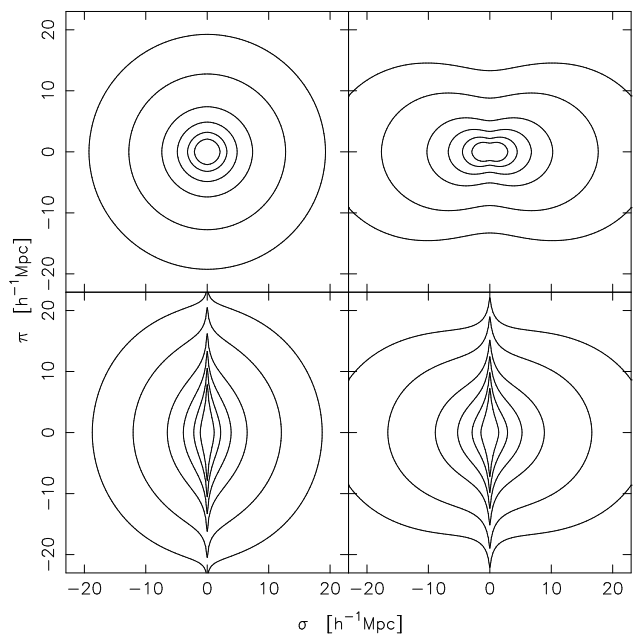}
\caption[]{Plot of model $\xisp$'s calculated as described in Section\
\ref{s:dis}. The lines represent contours of constant $\xisp$ = 4.0,
2.0, 1.0, 0.5, 0.2 and 0.1 for different models.  The top left panel
represents an undistorted correlation function ($a = 0, \beta = 0$),
the top right panel is a model with coherent infall added ($a = 0,
\beta = 0.4$), the bottom left panel is a model with just random
pairwise velocities added ($a = 500\kms, \beta = 0$) and the bottom
right panel has both infall and random motions added ($a = 500\kms$,
$\beta = 0.4$). These four models have $r_0 = 5.0\hmpc$ and $\gamma_r
= 1.7$.}
\label{f:modsp}
\end{figure}

To illustrate the effect of redshift-space distortions on the $\xisp$
plot we show four model $\xisp$'s in Fig.\ \ref{f:modsp}. If there
were no distortions, then the contours shown would be circular, as in
the top left panel due to the isotropy of the real-space correlation
function. On small $\sigma$ scales the random peculiar velocities
cause an elongation of the contours in the $\pi$ direction (the bottom
left panel). On larger scales there is the flattening of the contours
(top right panel) due to the coherent infall. The bottom right panel
is a model with both distortion effects included. Comparing the models
of $\xisp$ to the 2dFGRS measurements in Fig.\ \ref{f:sigpi} it is
clear that the data show the two distortion effects included in the
models. In Section\ \ref{s:gridfit} we use the data to constrain the
model directly, and deduce the best-fit model parameters.

\section{Estimating $\beta$}

Before using the model described above to measure the parameters
simultaneously, we first use methods that have been used in previous
studies. This allows a direct comparison between our results and
previous work.

\subsection{Ratio of $\xi$'s}

\begin{figure}
\psfig{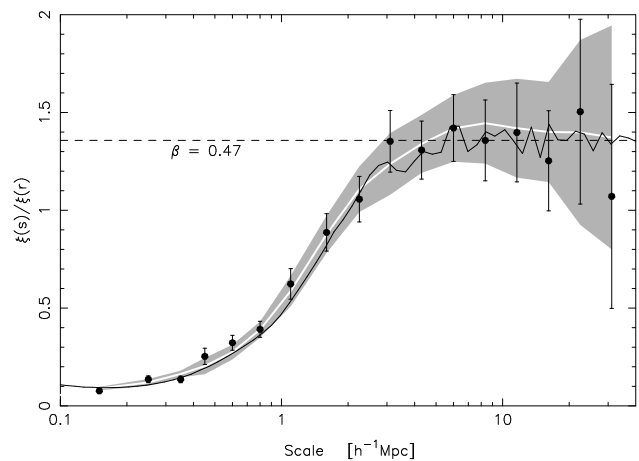}
\caption[]{The ratio of $\xis$ to $\xir$ for the 2dFGRS combined data
(solid points), and the Hubble Volume (solid line). The mean of the
mock catalogue results is also shown (white line), with the rms errors
shaded. The error bars on the 2dFGRS data are from the rms spread in
mock catalogue results.}
\label{f:ratios}
\end{figure}

The ratio of the redshift-space correlation function, $\xis$, to the
real-space correlation function, $\xir$, in the linear regime gives
an estimate of the redshift distortion parameter, $\beta$ (see
Eqn.\ \ref{e:xi0}),
\begin{equation}
\frac{\xis}{\xir} = 1 + \frac{2\beta}{3} + \frac{\beta^2}{5}.
\end{equation}
Our results for the combined 2dFGRS data, using the inverted form of
$\xir$, are shown in Fig.\ \ref{f:ratios} by the solid points. The
mean of the mock catalogue results is shown by the white line, with
the rms errors shaded and the estimate from the Hubble Volume is shown
by the solid line.  The data are consistent with a constant value, and
hence linear theory, on scales $\gtrsim 4\hmpc$.

The mock catalogues and Hubble Volume results asymptote to $\beta =
0.47$, the true value of $\beta$ in the mocks. The 2dFGRS data in the
range $8 - 30\hmpc$ are best-fit by a ratio of $\xirat$, corresponding
to $\beta = \betaxirat$.  The maximum scale that we can use in this
analysis is determined by the uncertainty on $\xir$ from the Saunders
\etal~inversion method discussed in Section\ \ref{s:inv}.

\begin{figure}
\psfig{figure=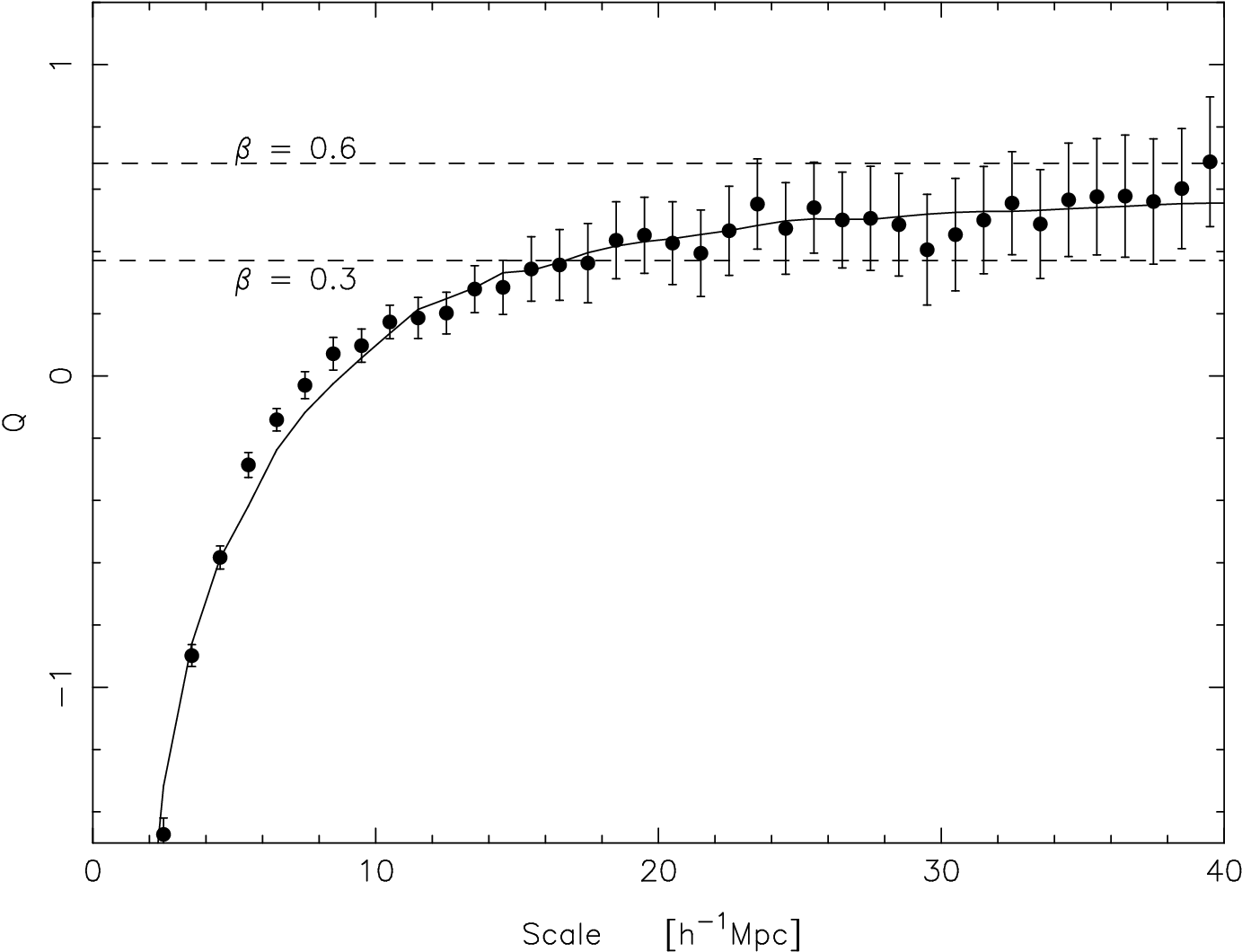,angle=-90,width=8.3cm,clip=}
\caption[]{The $Q$ factor for the combined 2dFGRS data, restricted to
scales not dominated by noise with error bars from the rms spread in
mock catalogue results. The two dashed lines show the expected answer
for different values of $\beta$ which approximate the $1\sigma$
errors. The solid line shows a model with $\beta = 0.49$ and $a =
506\kms$ (see Section\ \ref{s:fits}). The inset shows the result for
the NGP (solid points) and SGP (open points); the error bars are
placed alternately to avoid confusion.}
\label{f:Q}
\psfig{figure=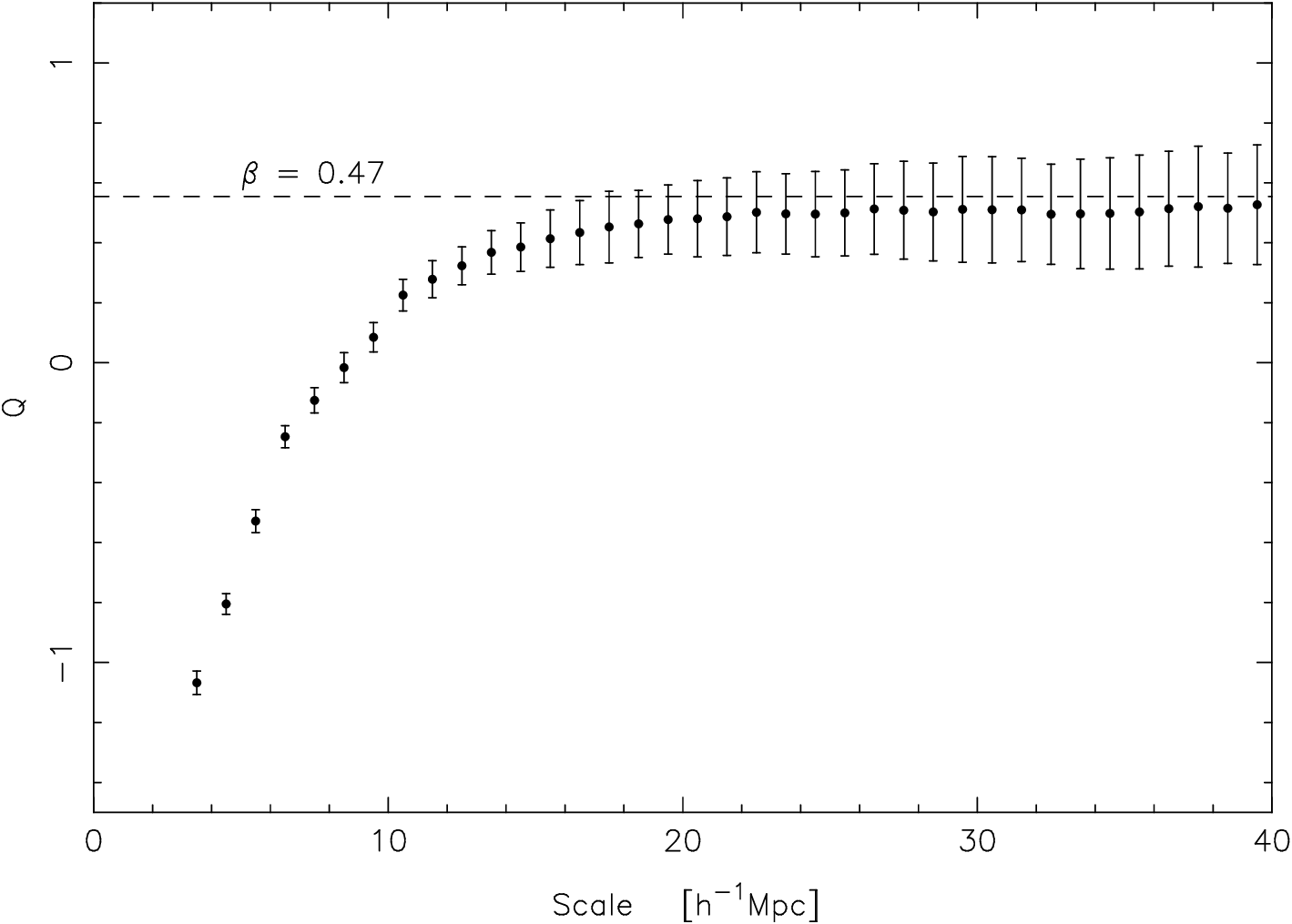,angle=-90,width=8.3cm,clip=}
\caption[]{The mean $Q$ factor for the mock catalogues with error bars
from the rms spread in mock catalogue results. The dashed line is the
true value of $\beta = 0.47$. The slight bias is caused by the
random peculiar velocities as discussed in the text.}
\label{f:mock_Q}
\end{figure}

\subsection{The quadrupole moment of $\xi$}
\label{s:Q}

We now measure $\beta$ using the quadrupole moment of the correlation
function (Hamilton 1992),
\begin{equation}
Q(s) = \frac{\frac{4}{3}\beta + \frac{4}{7}\beta^2}{1 + \frac{2}{3}\beta + \frac{1}{5}\beta^2} = \frac{\xi_2(s)}{\frac{3}{s^3}\int_{0}^{s}\xi_0(s'){s'}^{2}ds' - \xi_0(s)}
\end{equation}
where $\xi_{\ell}$ is given by,
\begin{equation}
\xi_{\ell}(s) = \frac{2\ell+1}{2}\int^{+1}_{-1}\xisp P_{\ell}(\ct)d\ct.
\end{equation}
These equations assume the random peculiar velocities are negligible
and hence measuring $Q$ gives an estimate of $\beta$. The random
uncertainties in this method are small enough that we obtain reliable
estimates on scales $< 40\hmpc$, as shown by the mock catalogues (see
below), but the data are noisy beyond these scales.

Fig.\ \ref{f:Q} shows $Q$ estimates for the combined 2dFGRS data with
the inset showing the NGP and SGP separately. The effect of the random
peculiar velocities can be clearly seen at small scales, causing $Q$
to be negative. The best-fit value to the combined data for $30 -
40\hmpc$ is $Q = \Qval$, which gives a value for $\beta = \betaQ$,
where the error is from the rms spread in the mock catalogue
results. The solid line represents a model with $\beta = 0.49$ and $a
= 506\kms$, which matches the data well (see Section\
\ref{s:fits}). Although asymptoting to a constant, the value of $Q$ in
the model is still increasing at $40\hmpc$. This shows that non-linear
effects do introduce a small systematic error even at these scales
though this bias is small compared to the random error.

To check whether this method can correctly determine $\beta$ we use
the mock catalogues. The data points in Fig.\ \ref{f:mock_Q} are the
mean values of $Q$ from the mock catalogues, with error bars on the
mean, and the dashed line is the true value of $\beta = 0.47$. The
data points seem to converge on large scales to the correct value of
$Q$. Fitting to each mock catalogue in turn for $30 - 40\hmpc$ gives a
mean $Q = 0.51 \pm 0.18$, corresponding to $\beta =
0.43^{+0.18}_{-0.16}$. As the models showed, the random velocities
will lead to an underestimate of $\beta$ even at $40\hmpc$, causing
the difference between the measured and true values. This all shows
that we can determine $\beta$ with a slight bias but the error bars
are large compared to the bias.

The $Q$ estimates from the individual mock catalogues show a high
degree of correlation between points on varying scales and so the
overall uncertainty in $Q$ from averaging over all scales $>30\hmpc$
is not much smaller than the uncertainty from a single point. It is
this fact which makes the spread in results from the mock catalogues
vital in the estimation of the errors on our result (also see Section\
\ref{s:john}).

\section{The peculiar velocity distribution}
\label{s:expvel}

To this point we have assumed that the random peculiar velocity
distribution has an exponential form (Eqn.\ \ref{e:fv}). This form has
been used by many authors in the past and has been found to fit the
data better than other forms (e.g. Ratcliffe \etal~1998). We test this
for the 2dFGRS data by following a method similar to that of Landy,
Szalay and Broadhurst (1998, hereafter LSB98). To extract the peculiar
velocity distribution, we need to deconvolve the real-space correlation
function from the peculiar velocity distribution.

\subsection{The method}

We first take the 2-d Fourier transform of the $\xisp$ grid to give
$\hat \xi(k_\sigma, k_\pi)$ and then take cuts along the $k_\sigma$
and $k_\pi$ axes which we denote by $\Sigma(k)$ and $\Pi(k)$
respectively, so $\Sigma(k) = \hat \xi(k_\sigma=k, k_\pi=0)$ and
$\Pi(k) = \hat \xi(k_\sigma=0, k_\pi=k)$.  By the slicing-projection
theorem (see LSB98) these cuts are equivalent to the Fourier
transforms of the real-space projections of $\xisp$ onto the $\sigma$
and $\pi$ axes.  The projection of $\xisp$ onto the $\sigma$ axis is a
distortion free measurement of $\Xi$ but the projection onto the $\pi$
axis gives us $\Xi$ convolved with the peculiar velocity distribution,
ignoring the effects of large-scale bulk flows. Since a convolution in
real space is a multiplication in Fourier space, the ratio of
$\Sigma(k)$ to $\Pi(k)$ is the Fourier transform, ${\mathcal
F}[f(v)]$, of the velocity distribution that we want to estimate. All
that is left is to inverse Fourier transform this ratio to obtain the
peculiar velocity distribution, $f(v)$. LSB98 cut their dataset at
$32\hmpc$ and applied a Hann smoothing window; we use all the raw
data. Landy (2002, hereafter L02) used the LSB98 method on the 100k
2dFGRS Public Release data and his results are discussed below.

Fitting an exponential to the resulting $f(v)$ curve gives a value for
$a$ assuming that the infall contribution to the velocity distribution
is negligible. LSB98 and L02 claim that their method is not sensitive
to the infall velocities. We show here that this is not the case. The
additional structure in the Fourier transform of the velocity
distribution found by L02 is a direct consequence of the infall
velocities.

\subsection{Testing the models}
\label{s:vfv}

\begin{figure}
\psfig{figure=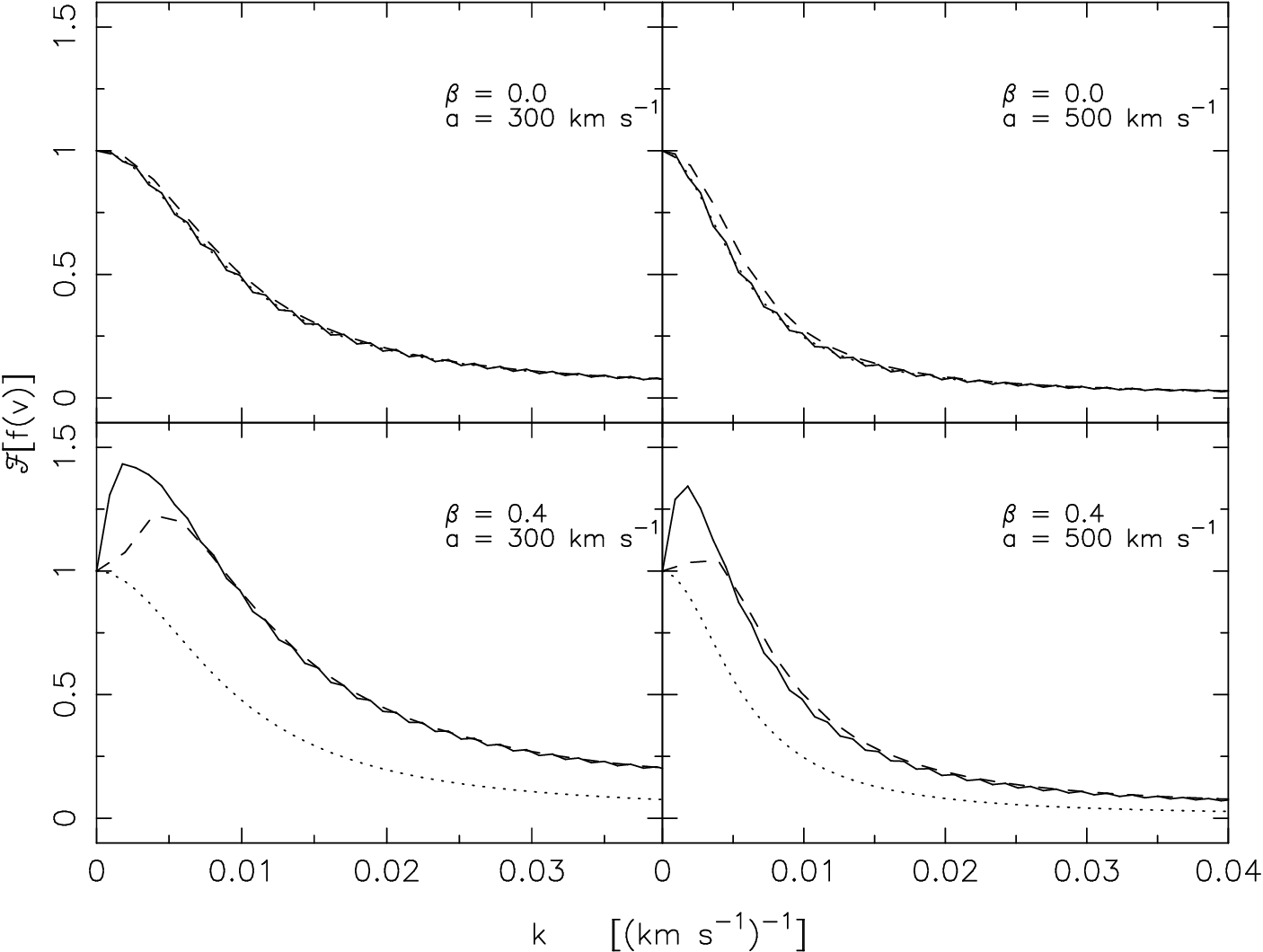,angle=-90,width=8.3cm,clip=}
\caption[]{The Fourier transform of the peculiar velocity distribution
for various parameters (see labels). The solid line is for a model
with no smoothing and using all scales $< 70\hmpc$. The dashed line is
for a model cut at $32\hmpc$ and smoothed with a Hann window (like
L02). The dotted line is the Lorentzian equivalent of the input
exponential peculiar velocity distribution (this coincides with the
solid line in the top panels).}
\label{f:fft}
\psfig{figure=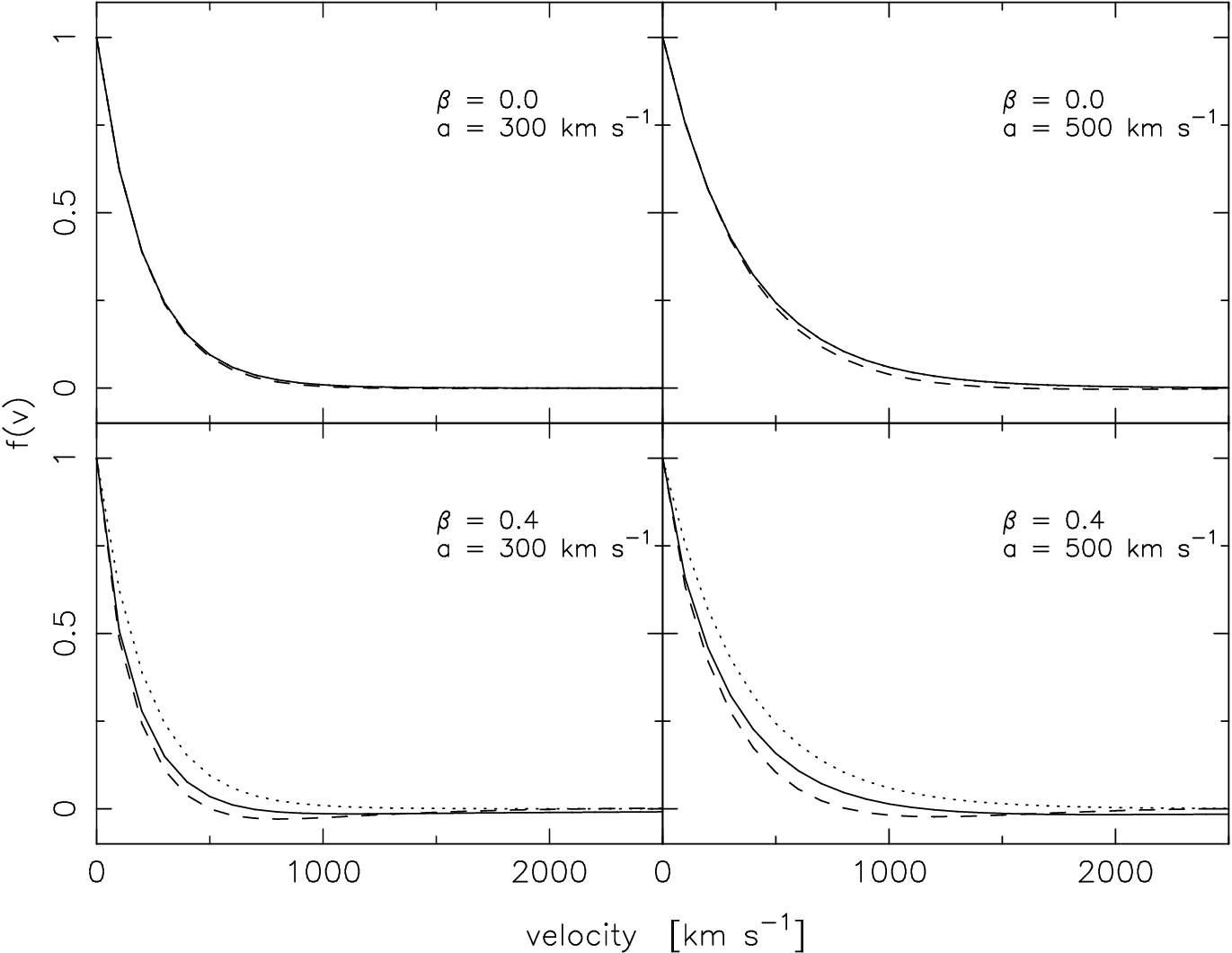,angle=-90,width=8.3cm,clip=}
\caption[]{The peculiar velocity distribution for various models (see
labels). The solid line is the recovered distribution for a model with
no smoothing and using all scales $< 70\hmpc$. The dashed line is the
recovered distribution for a model cut at $32\hmpc$ and smoothed with
a Hann window. The dotted line is the input exponential peculiar
velocity distribution (this coincides with the solid line in the top
panels).}
\label{f:fv}
\end{figure}
\begin{figure}
\psfig{figure=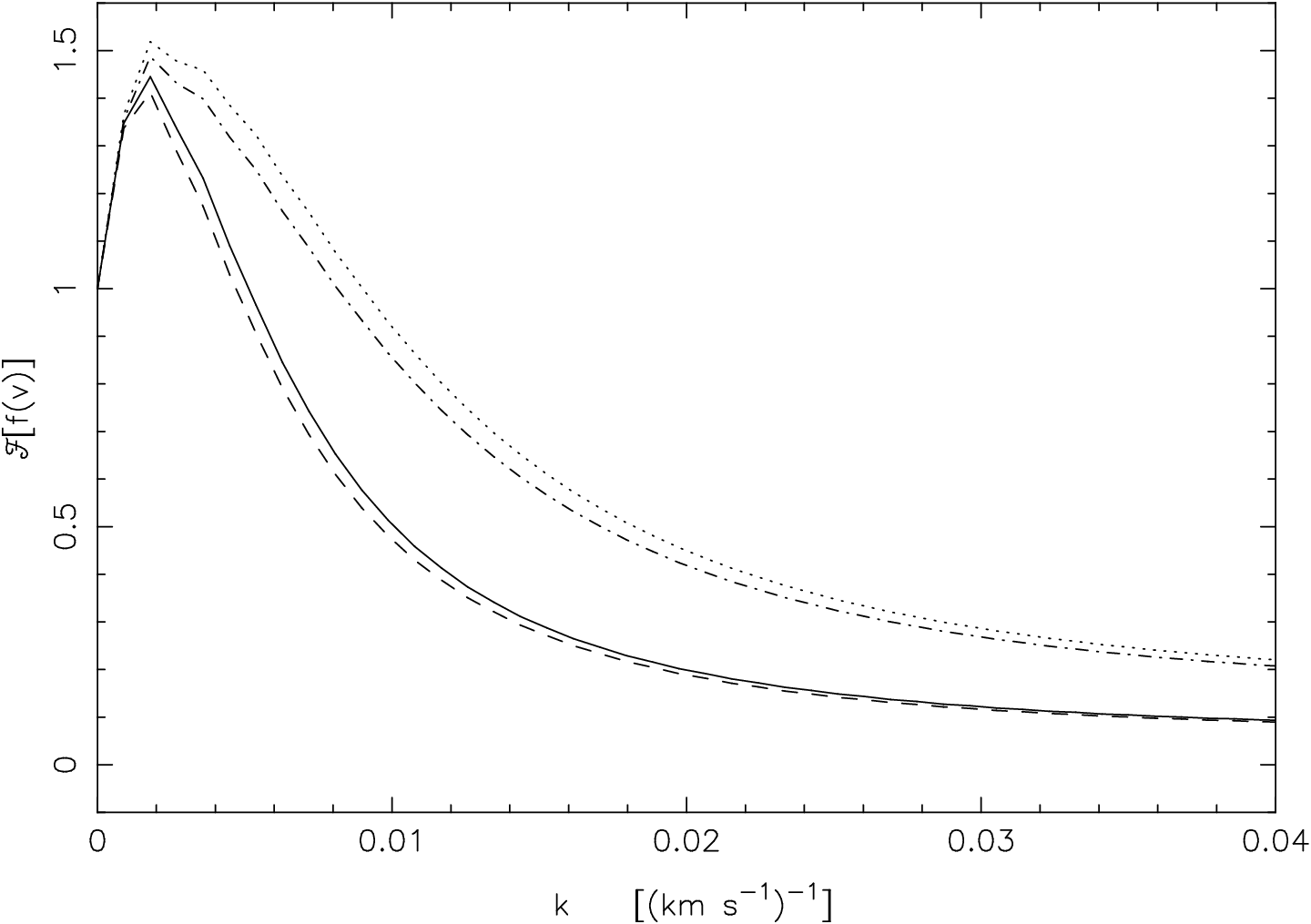,angle=-90,width=8.3cm,clip=}
\caption[]{The Fourier transform of the peculiar velocity distribution
for a model with $a = 500\kms$ (dashed line), $a = 300\kms$ (dotted
line). Also shown is a model with $a$ decreasing from $500\kms$ to
$300\kms$ from $\sigma = 0$ to $\sigma = 20\hmpc$ (solid line) and a
model with $a$ increasing from $300\kms$ to $500\kms$ from $\sigma =
0$ to $\sigma = 20\hmpc$ (dot-dashed line). $\beta = 0.5$ for all four
models.}
\label{f:vfv}
\end{figure}

To test the LSB98 method we apply the technique to our models,
described in Section\ \ref{s:dis}, with and without a $\beta=0.4$
infall factor, using various scales, and with and without a Hann
window. In Fig.\ \ref{f:fft} we show the Fourier transform of the
peculiar velocity distribution and in Fig.\ \ref{f:fv} we show the
distribution function itself.

It is clear from Fig.\ \ref{f:fft} that the shape of the Fourier
transform at small $k$ is quite badly distorted by the infall
velocities.  This leads to a systematic error in the actual velocity
distribution as seen in Fig.\ \ref{f:fv}, where the measured peculiar
velocity dispersions are biased low, especially in the case where a
smoothing window and a limited range of scales are used. In particular
the peak of the Fourier transform is not at $k=0$, and the inferred
$f(v)$ goes negative for a range of velocities (dashed lines in the
lower panels of Fig.\ \ref{f:fv}).  This clearly cannot be interpreted
as a physical velocity distribution; the method infers negative values
because the input model $\xisp$ is not consistent with the initial
assumption of the method, which is that all of the distortion in
$\xisp$ is due to random peculiar velocities. We conclude that both
types of peculiar velocity need to be considered when making these
measurements, and so our preferred results come from directly fitting
to $\xisp$.

A further complication with the real data is that $f(v)$ may depend on
the pair separation (see discussion in Section 7.3).  The solid line
in Fig.\ \ref{f:vfv} shows ${\mathcal F}[f(v)]$ for a model where $a$
varies from 500$\kms$ at $\sigma =0 $ to 300$\kms$ at $\sigma =
20\hmpc$. This is compared to a model with $a=500\kms$ (dashed line),
a model with $a=300\kms$ (dotted line) and a model where $a$ varies
from 300$\kms$ at $\sigma =0$ to $500\kms$ at $\sigma =20\hmpc$
(dot-dashed line). The models with varying $a$ are very close to their
respective constant $a$ models at all $k$ values, showing that this
method leads to an estimate of ${\mathcal F}[f(v)]$ determined mainly
by the value of $a$ at small $\sigma$.

\begin{figure}
\psfig{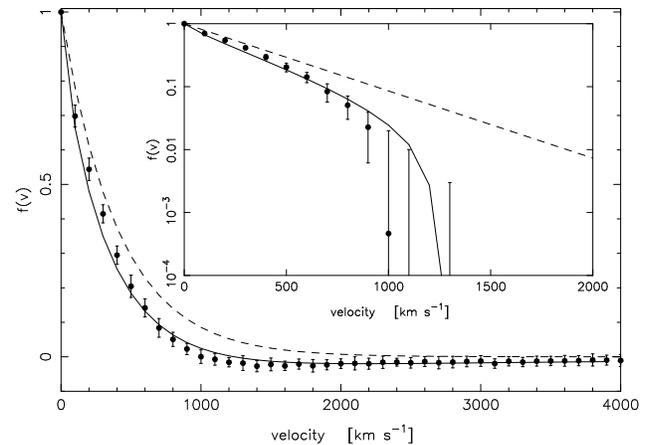}
\caption{The recovered velocity distribution for the mock
catalogues. Filled points are the mean result with error bars from the
scatter between catalogues. This is compared to a pure exponential
distribution with $a = 575\kms$ (dashed line) and a model with $a =
575\kms$ and $\beta=0.47$ (solid line).}
\label{f:mveld}
\end{figure}
\begin{figure}
\vspace*{0.2cm}
\psfig{figure=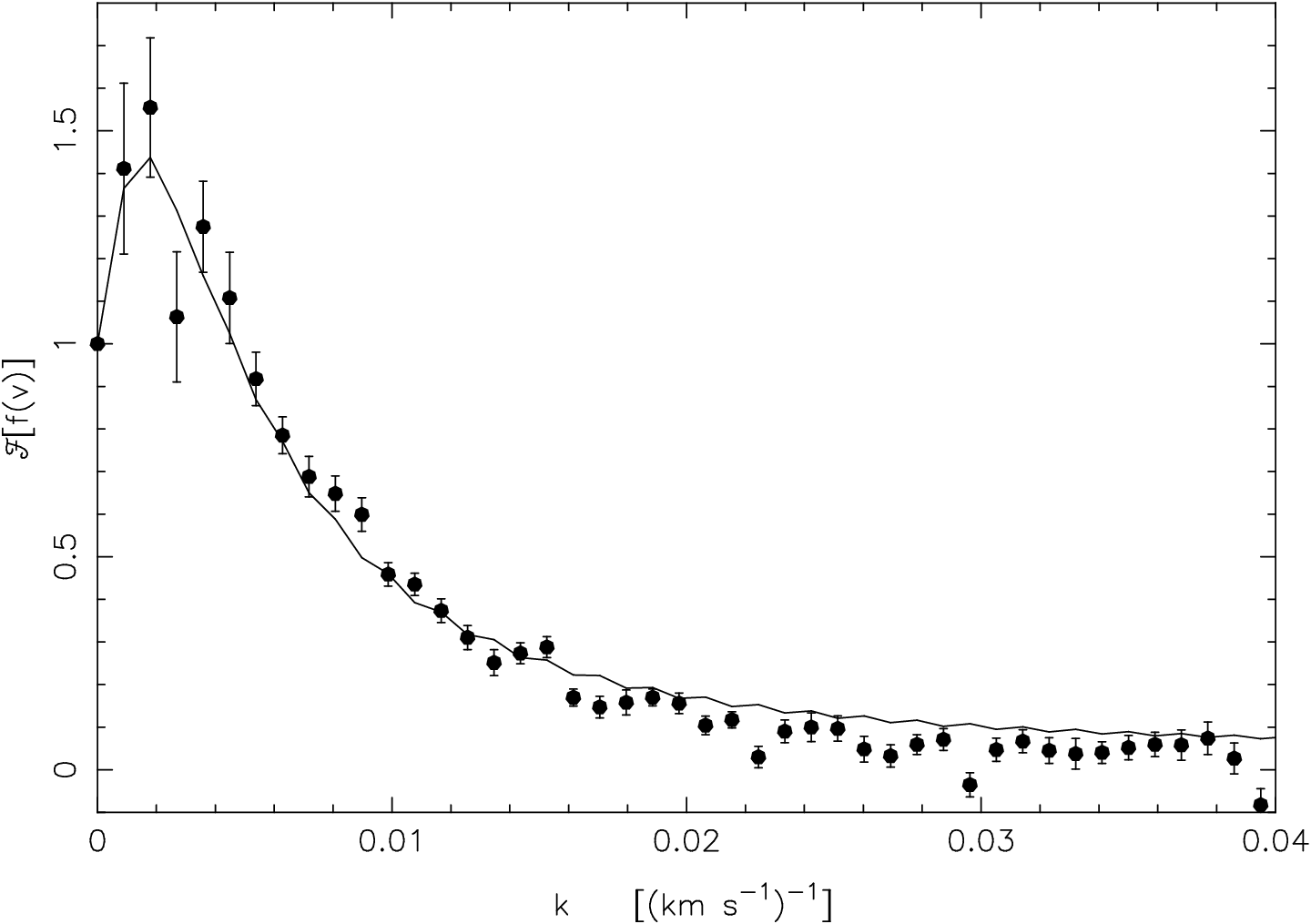,angle=-90,width=8.3cm,clip=}
\caption[]{The Fourier transform of the peculiar velocity distribution
for the combined 2dFGRS data (solid points) compared to a model
distribution with $a = 570\kms$ and $\beta = 0.49$ (solid line).}
\label{f:ffta}
\vspace*{0.2cm}
\psfig{figure=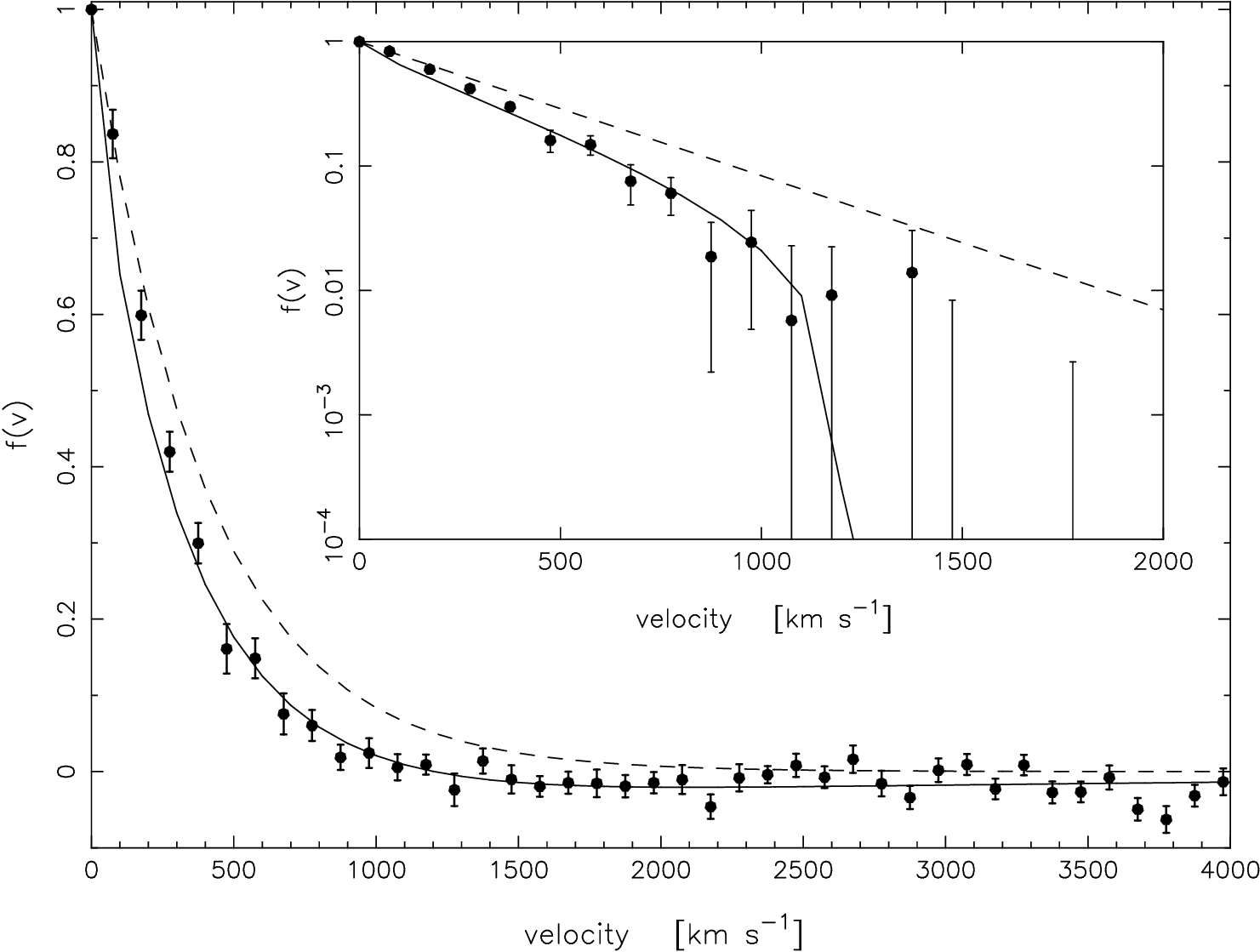,angle=-90,width=8.3cm,clip=}
\caption[]{The combined 2dFGRS peculiar velocity distribution (solid
points), compared to a pure exponential distribution with $a =
570\kms$ (dashed line) and a model with $a = 570\kms$ and $\beta =
0.49$ (solid line). The error bars are from the scatter in mock
results.}
\label{f:velda}
\end{figure}

\subsection{The mock catalogues}

The mean of the peculiar velocity distributions for the mock
catalogues is shown in Fig.\ \ref{f:mveld}. The distribution is
compared to a model, shown as the solid line, with $\beta = 0.47$ and
an exponential $f(v)$, with dispersion, $a = 575\kms$. The exact form
of the peculiar velocities in the Hubble Volume, and hence mock
catalogues, is not explicitly specified and it should not be expected
to conform to this model exactly.

\subsection{The 2dFGRS data}

The Fourier transform of the peculiar velocity distribution for the
combined 2dFGRS data are shown in Fig.\ \ref{f:ffta} compared to a
best-fit model with $\beta = 0.49 \pm 0.05$ and $a = 570 \pm
25\kms$. Fig.\ \ref{f:velda} shows the peculiar velocity distribution
itself compared to the same model. We showed in Section\ \ref{s:vfv}
(with Fig.\ \ref{f:vfv}) that this was likely to be the value of $a$
at small $\sigma$. The distribution of random pairwise velocities does
appear to have an exponential form, with a $\beta$ influence. Sheth
(1996) and Diaferio \& Geller (1996) have shown that an exponential
peculiar velocity distribution is a result of gravitational processes.

Ignoring the infall L02 found $a = 331\kms$, using the smaller,
publicly available, sample of 2dFGRS galaxies. We made the same
approximations and repeated his procedure on our larger sample, and
find $a = 370\kms$. Using our data grid out to $70\hmpc$, with no
smoothing and ignoring $\beta$, gives $a = 457\kms$. We have shown
that the result in L02 is biased low by ignoring $\beta$ and that the
infall must be properly considered in these analyses. As shown in
Fig.\ \ref{f:velda}, our data are reasonably well described by an
exponential model with $\beta = 0.49$ and $a = 570\kms$.

\section{Fitting to the $\xisp$ grid}
\label{s:gridfit}

\subsection{Results}
\label{s:fits}

We now fit our $\xisp$ data grid to the models described in Section
\ref{s:dis}, assuming a power-law form for the real-space correlation
function. This model has four free parameters, $\beta$, $r_0$,
$\gamma_r$ and $a$. The fits to the data are done by minimising
\begin{equation}
E = \sum\left(\frac{\log[1+\xi]_{\rm{model}} - \log[1+\xi]_{\rm{data}}}{\log[1+\xi+\delta\xi]_{\rm{data}}-\log[1+\xi-\delta\xi]_{\rm{data}}}\right)^2,
\end{equation}
for $s < 20\hmpc$, where $\delta\xi$ is the rms of $\xi$ from the mock
catalogues for a particular $\sigma$ and $\pi$. This is like a simple
$\chi^2$ minimization, but the points are not independent. We tried a
fit to $\xi$ directly but found that it gave too much weight to the
central regions and so instead we fit to $\log[1+\xi]$ so that the
overall shape of the contours has an increased influence on the
fit. The best-fit model parameters are listed in Table\
\ref{t:sp2}. The errors we quote are the rms spread in errors from
fitting each mock catalogue in the same way.

There are two key assumptions made in the construction of these
models. Firstly, although the contours match well at small scales,
there are good reasons to believe that our linear theory model will
not hold in the non-linear regime for $s \lesssim 8\hmpc$. Secondly,
we have assumed the power-law model for $\xir$ and we have seen
evidence that this is not completely realistic. Using non-power law
forms will also allow us to probe to larger scales.

To test whether our result is robust to these assumptions we firstly
reject the non-linear regime corresponding to $s < 8\hmpc$. Then, we
use the shape of the Hubble Volume $\xir$ instead of a power-law, and
finally we extend the maximum scale to $s = 30\hmpc$. We showed in
Section\ \ref{s:inv} that the Hubble Volume shape gives a good match
to the data over the range $8 < s < 30\hmpc$ (the Appendix gives the
relevant equations for performing the $\beta$ infall calculation
without a power-law assumption).

\begin{figure}
\psfig{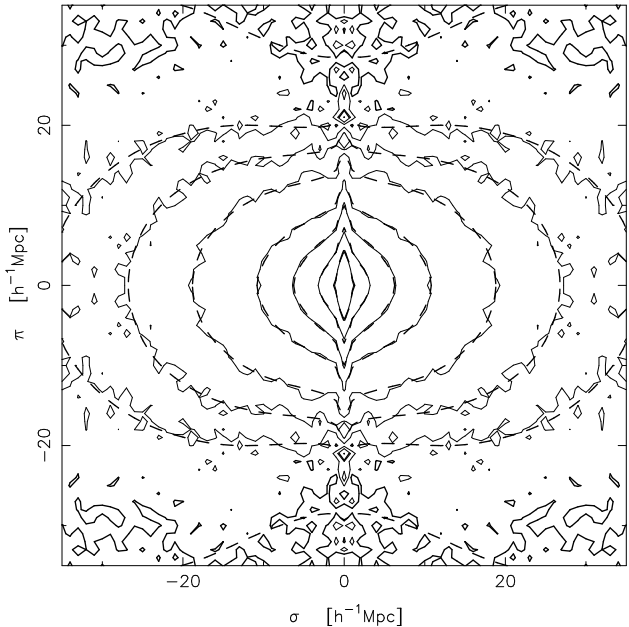}
\caption[]{Contours of $\xisp$ for the 2dFGRS combined data (solid
lines) and the best-fit model (see Table\ \ref{t:sp2}) using the
Hubble Volume $\xir$ fitted to scales $8 < s < 30\hmpc$ (dashed
lines). Contour levels are at $\xi$ = 4.0, 2.0, 1.0, 0.5, 0.2, 0.1,
0.05 and 0.0 (thick line).}
\label{f:sps2}
\end{figure}

We find that the best-fit parameters change very little with these
changes but when using the Hubble Volume $\xir$, the quality of the
fit improves significantly. The best-fit model is compared to the data
in Fig.\ \ref{f:sps2}. Notice the excellent agreement on small scales
even though they are ignored in the fitting process. The best-fit
parameters are listed in Table\ \ref{t:sp2}, and we adopt these
results as our final best estimates finding $\beta = \betaoverall$.

If we repeat our analysis on the mock catalogues we find a mean value
of $\beta = 0.475 \pm 0.090$ (cf. the expected value of $\beta =
0.47$, Section\ \ref{s:mocks}), showing that we can correctly
determine $\beta$ using this type of fit. When fitting the mock
catalogues it became clear that $\beta$ and $a$ are correlated in this
fitting procedure, as we have seen already with other methods. We use
the mock catalogues to measure the linear correlation coefficient, $r$
(Press \etal~1992), which quantifies this correlation, and find that,
between $\beta$ and $a$, $r = 0.66$. If we knew either parameter
exactly, the error on the other would be smaller than quoted.

We also tried other analytical forms for the correlation function and
also different scale limits and found that some combinations shifted
the results by $\sim 1\sigma$.

\begin{table}
\caption{Best-fit parameters to the $\xisp$ grids with errors from the
rms spread in mock catalogue results.}
\label{t:sp2}
\begin{tabular}{llll}
\hline
Parameter     & SGP                & NGP               & Combined\\
\hline\hline
\multicolumn{4}{l}{Power law $\xir$:~($0 < s < 20\hmpc$)}\\
\hline
$\beta$       & $0.53 \pm 0.06$  & $0.48 \pm 0.08$ & $0.51 \pm 0.05$\\
$r_0~(\hmpcb)$& $5.63 \pm 0.26$  & $5.52 \pm 0.29$ & $5.58 \pm 0.19$\\
$\gamma_r$    & $1.66 \pm 0.06$  & $1.76 \pm 0.07$ & $1.72 \pm 0.05$\\
$a~(\kmsn)$   & $497 \pm 24$     & $543 \pm 26$    & $522 \pm 16$\\
\hline\hline
\multicolumn{4}{l}{Power law $\xir$:~($8 < s < 20\hmpc$)}\\
\hline
$\beta$       & $0.45 \pm 0.10$  & $0.35 \pm 0.12$ & $0.49 \pm 0.09$\\
$r_0~(\hmpcb)$& $6.03 \pm 0.36$  & $6.06 \pm 0.41$ & $5.80 \pm 0.25$\\
$\gamma_r$    & $1.74 \pm 0.08$  & $1.88 \pm 0.10$ & $1.78 \pm 0.06$\\
$a~(\kmsn)$   & $457  \pm 49$    & $451  \pm 51$   & $514  \pm 31$\\
\hline\hline
\multicolumn{4}{l}{Hubble Volume $\xir$:~($8 < s < 20\hmpc$)}\\
\hline
$\beta$       & $0.47 \pm 0.12$  & $0.50 \pm 0.14$ & $0.49 \pm 0.10$\\
$a~(\kmsn)$   & $446 \pm 73$     & $544 \pm 67$    & $495 \pm 46$\\
\hline\hline
\multicolumn{4}{l}{Hubble Volume $\xir$:~($8 < s < 30\hmpc$)}\\
\hline
$\beta$       & $0.48 \pm 0.11$ & $0.47 \pm 0.13$ & $\betaoverall$\\
$a~(\kmsn)$   & $450 \pm 81$    & $545 \pm 85$    & $\aoverall$\\
\hline
\end{tabular}
\end{table}

\subsection{Comparison of methods}

We have now estimated the real-space clustering parameters using three
different methods. In Section\ \ref{s:xir_comp}, we saw that the
projection and inversion methods gave essentially identical results
for $r_0$ and $\gamma_r$ whereas using 2-d fits we get slightly higher
values for $r_0$.

If $\xir$ was a perfect power-law the different methods would give
unbiased results for the parameters, but we have seen evidence that
this assumption is not true. The methods therefore, give different
answers as a result of the different scales and weighting schemes
used, as well as the vastly different treatments of the redshift-space
distortions.

\subsection{Previous 2dFGRS results}
\label{s:john}

It is worth contrasting our present results with those obtained in a
previous 2dFGRS analysis (Peacock et al. 2001). This was based on the
data available up to the end of 2000: a total of 141\,402
redshifts. The chosen redshift limit was $z_{\rm max}=0.25$, yielding
127\,081 galaxies for the analysis of $\xi(\sigma,\pi)$. The present
analysis uses 165\,659 galaxies, but to a maximum redshift of 0.2.
Because galaxies are given a redshift-dependent weight, this
difference in redshift limit has a substantial effect on the volume
sampled. For a given area of sky, changing the redshift limit from
$z_{\rm max}=0.2$ to $z_{\rm max}=0.25$ changes the total number of
galaxies by a factor of only 1.08, whereas the total comoving volume
within $z_{\rm max}$ increases by a factor of 2. Allowing for the
redshift-dependent weight used in practice, the difference in
effective comoving volume for a given area of sky due to the variation
in redshift limits becomes a factor of 1.6. Since the effective area
covered by the present data is greater by a factor of $165\,659 /
(127\,081 / 1.08) = 1.4$, the total effective comoving volume probed
in the current analysis is in fact 15\% smaller than in the 2001
analysis; this would suggest random errors on clustering statistics
about 7\% larger than previously. Of course, the lower redshift limit
has several important advantages: uncertainties in the selection
function in the tail of the luminosity function are not an issue (see
Norberg et al. 2002a); also, the mean epoch of measurement is closer
to $z=0$. Given that the sky coverage is now more uniform, and that
the survey mask and selection function have been studied in greater
detail, the present results should be much more robust.

The other main difference between the present work and that of Peacock
\etal~(2001) lies in the method of analysis. The earlier work
quantified the flattening of the contours of $\xi(\sigma,\pi)$ via the
quadrupole-to-monopole ratio, $\xi_2(s)/\xi_0(s)$. This is not to be
confused with the quantity $Q(s)$ from Section\ \ref{s:Q}, which uses
an integrated clustering measure instead of $\xi_0(s)$. This is
inevitably more noisy, as reflected in the error bar,
$\delta\beta=0.17$, resulting from that method. The disadvantage of
using $\xi_2(s)/\xi_0(s)$ directly, however, is that the ratio depends
on the true shape of $\xi(r)$. In Peacock et al. (2001), this was
assumed to be known from the de-projection of angular clustering in
the APM survey (Baugh \& Efstathiou 1993); in the present paper we
have made a detailed internal estimate of $\xi(r)$, and considered the
effect of uncertainties in this quantity. Apart from this difference,
the previous method of fitting to $\xi_2(s)/\xi_0(s)$ should, in
principle, give results that are similar to our full fit to
$\xi(\sigma,\pi)$ in Section\ \ref{s:fits}. The key issue in both
cases is the treatment of the errors, which are estimated in a fully
realistic fashion in the present paper using mock samples. The
previous analysis used two simpler methods: an empirical error on
$\xi_2(s)/\xi_0(s)$ was deduced from the NGP--SGP difference, and
correlated data were allowed for by estimating the true number of
degrees of freedom from the value of $\chi^2$ for the best-fit
model. This estimate was compared with a covariance matrix built from
multiple realizations of $\xi(\sigma,\pi)$ using Gaussian fields;
consistent errors were obtained. We applied the simple method of
Peacock \etal~(2001) to the current data, keeping the assumed APM
$\xi(r)$, and obtained the marginalized result $\beta=0.55 \pm
0.075$. The comparison with our best estimate of $\beta=0.49 \pm 0.09$
indicates that the systematic errors in the previous analysis (from
e.g. the assumed $\xi[r]$) were not important, but that the previous
error bars were optimistic by about 20\%.

\subsection{Peculiar velocities as a function of scale}
\label{s:pecvels}

\begin{figure*}
\psfig{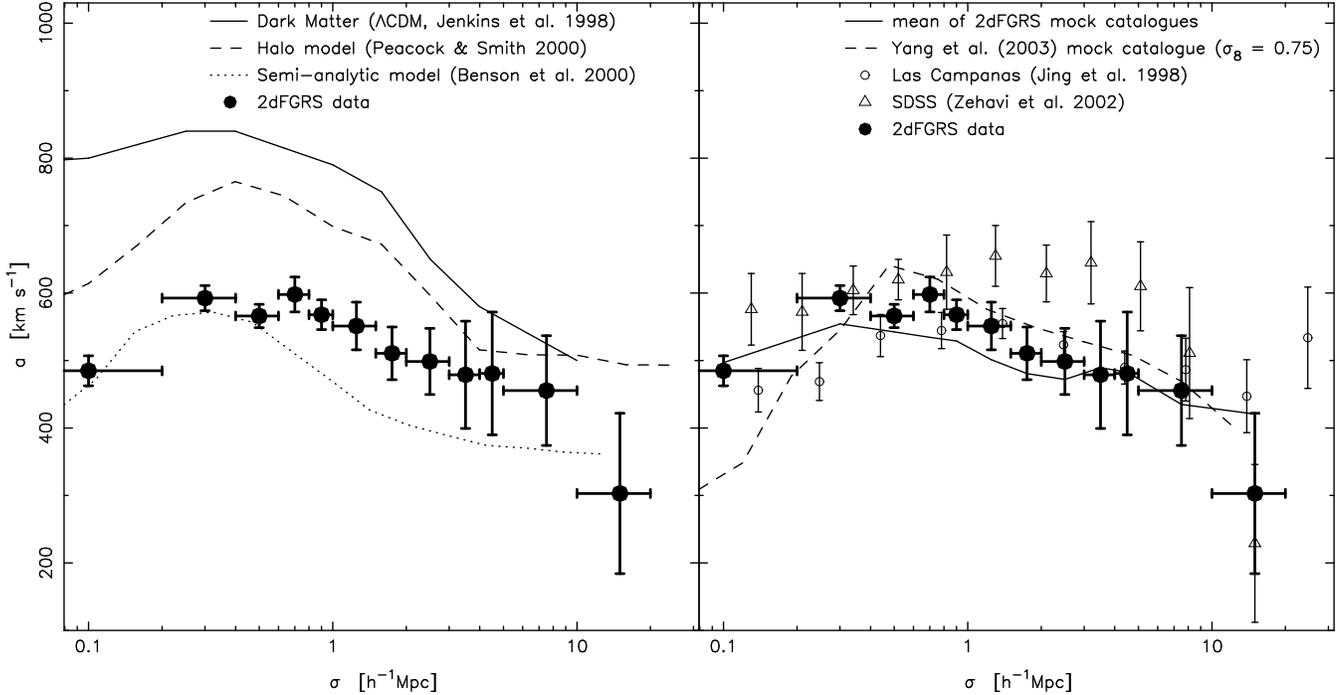}
\caption[]{The variation of $a$ with projected separation,
$\sigma$. Left panel: 2dFGRS data compared to some analytical models,
as indicated in the legend. Right panel: 2dFGRS data compared to
other redshift surveys and simulated catalogues, as indicated in the
legend. The 2dFGRS data and simulated catalogue results use $\beta$
to calculate the infall velocities whereas the other results assume a
functional form (see discussion in Section\ \ref{s:pecvels}).}
\label{f:va}
\end{figure*}

There has been much discussion in the literature on whether or not the
pairwise peculiar velocity dispersion, $a$, is a function of projected
separation, $\sigma$. Many authors have used $N$-body simulations to
make predictions for what might be observed. Davis \etal~(1985) found
that the pairwise velocity dispersion of cold dark matter remains
approximately constant on small scales, decreases by about 20-30\% on
intermediate scales and is approximately constant again on large
scales. Cen, Bahcall \& Gramann (1994) found a similar overall
behaviour as did Jenkins \etal~(1998) whose results are plotted in the
left panel of Fig.\ \ref{f:va} as the solid line for a $\lcdm$
cosmology. The dashed line is from Peacock \& Smith (2000) who used
the halo model to predict the peculiar velocities for the galaxy
distribution. Kauffmann \etal~(1999) and Benson \etal~(2000) used the
GIF simulations combined with semi-analytic models of galaxy formation
and the galaxy predictions of Benson \etal~(2000) are shown by the
dotted line. These predictions generally assume $\sigma_8 = 0.9$, but
there is evidence that $\sigma_8$ could be $10\%$ lower than this
(Spergel \etal~2003) and so the pairwise velocity dispersions implied
would also be lower.

Observationally, Jing \etal~(1998) measured the pairwise velocity
dispersion in the Las Campanas Redshift Survey and found no
significant variation with scale. We note again that the errors for
the LCRS ignore the effects of cosmic variance and are likely to be
underestimates. Zehavi \etal~(2002) used the SDSS data and found that
$a$ decreased with scale for $\sigma \gtrsim 5\hmpc$. These
observations are plotted in the right panel of Fig.\ \ref{f:va}. All
these observations have assumed a functional form for the infall
velocities (or `streaming') and not used $\beta$ directly. We have
already shown that proper consideration of the infall parameter is
vital in such studies. Indeed, Zehavi \etal\ (2002) say that their
estimates of $a$ for $\sigma > 3\hmpc$ depend significantly on their
choice of streaming model. This factor, along with a dependence of $a$
on luminosity and galaxy type may help to explain the differences
between the 2dFGRS and SDSS results.

The difference in results from Section\ \ref{s:vfv} which measured the
value of $a$ at small $\sigma$ ($570\kms$), and from using the $\xisp$
grid ($506\kms$), which measures an average value, hints that there
may be such an dependence of $a$ on $\sigma$ in the 2dFGRS data. We
test for variations in $a$ by repeating the fits described in Section\
\ref{s:fits} using a global $\beta, r_0$ and $\gamma_r$ but allowing
$a$ to vary in each $\sigma$ slice. The results are shown in Fig.\
\ref{f:va}, compared with the results from other surveys, and
numerical simulations as discussed above. The value of $506\kms$
obtained from the 2-d fit for scales $> 8\hmpc$ is close to the value
at $8\hmpc$ where most of the signal is coming from. The value of
$570\kms$ obtained from the Fourier transform technique agrees well
with the results found for $\sigma < 1\hmpc$. The values of $\beta,
r_0$ and $\gamma_r$ are essentially unchanged when fitting in this
way. We note again that the effects of the infall must be properly
taken into account in these measurements. We also note that we used
our linear, power-law model on all scales, but we have seen that this
is a reasonable approximation on non-linear scales.

We see that the overall shape of the 2dFGRS results are fairly
consistent with, though slightly flatter than the semi-analytic
predictions, but the amplitude is certainly a little different, which
could be due to the value of $\sigma_8$ used in the models, as
discussed above. We also plot the mean of the mock catalogue results
(solid line), and the results of a simulated catalogue (dashed line)
of Yang \etal\ (2003, with $\sigma_8 = 0.75$) and these match the real
data well.

\section{Constraining $\boldsymbol{\om}$}

We take the value of $\beta$ measured from the multi-parameter best
fit to $\xisp$,
\begin{equation}
\beta(L_s, z_s) = \betaoverall,
\end{equation}
which is measured at the effective luminosity, $L_s$, and redshift,
$z_s$, of our survey sample. In Section\ \ref{s:2dfdata} we quoted
these values, which are the applicable mean values when using the
$J_3$ weighting and redshift cuts employed, as $L_s \approx
1.4L^{\ast}$ and $z_s \approx 0.15$. We also note here that if we
adopt an $\om = 1$ geometry we find that $\beta = 0.55$, within the
quoted 1$\sigma$ errors.

\subsection{Redshift effects}
\label{s:zeffect}

The redshift distortion parameter can be written as,
\begin{equation}
\beta = \frac{f(\om, \oml, z)}{b}.
\end{equation}
where $f=d\ln D/d\ln a$, and $D$ is the linear fluctuation growth
factor and $a$ is the expansion factor. A good approximation for $f$,
at all $z$, in a flat Universe, was given by Lahav \etal~(1991),
\begin{equation}
f = \om^{0.6} + (2 - \om - \om^2)/140 \approx \om^{0.6},
\end{equation}
and so to constrain $\om$ from these results we need an estimate of
$b$. There have been two recent papers describing such measurements.

Verde \etal\ (2002) measured $b(L_s, z_s)$ from an analysis of the
bispectrum of 2dFGRS galaxies. Their results depend strongly on the
pairwise peculiar velocity dispersion, $a$, assumed in their analysis.
They used the result of Peacock \etal\ (2001), who found $a = 385
\kms$, somewhat lower than our new value of $\approx 500 \kms$. To
derive $\om$ using these results would not therefore be consistent and
so a new bispectrum analysis is in preparation.

\begin{figure}
\psfig{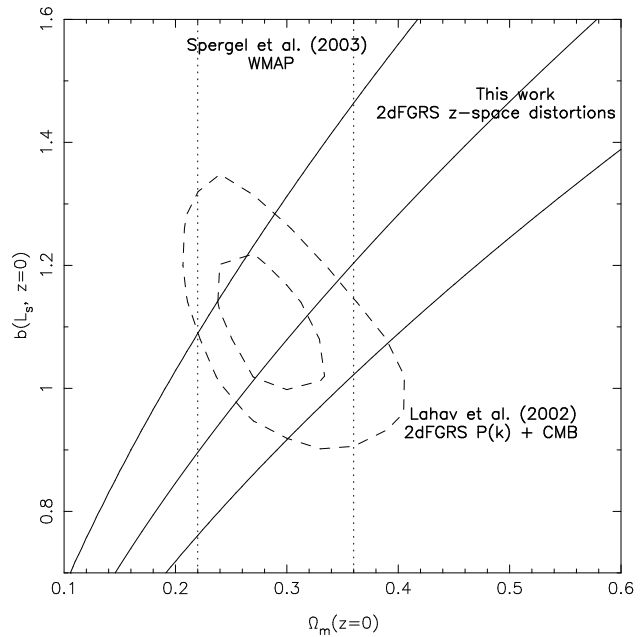}
\caption[]{Constraints on $\om$ and $b$ - Solid lines: Best-fit and
$1\sigma$ error contours on $\beta$ from this work, evolved to the
present day (see Section\ \ref{s:zeffect}). Dashed lines: $1\sigma$ and
$2\sigma$ error contours from Lahav \etal~(2002). Dotted lines:
$1\sigma$ constraints from {\it WMAP} (Spergel \etal~2003).}
\label{f:omvb}
\end{figure}

Lahav \etal\ (2002) combined the estimate of the 2dF power spectrum,
$P(k)$ (Percival \etal\ 2001), with pre-{\it WMAP} results from the
CMB to obtain an estimate of $b$, but this value is also dependent on
$\om$. Their likelihood contours\footnote{The bias parameter measured
by Lahav \etal\ (2002) depends on $\tau$, the optical depth of
reionisation, as $b \propto \exp(-\tau)$. The plotted results do not
include this effect, which could be significant, and this is discussed
further in Section\ \ref{s:omcomp}.} are reproduced in Fig.\
\ref{f:omvb}, as the dashed lines. They also introduced a `constant
galaxy clustering' model for the evolution of $b$ with $z$. Following
these equations we can evolve our measured $\beta$ to the present day
and estimate
\begin{equation}
\beta(L_s, z=0) = 0.45 \pm 0.08
\end{equation}
and these contours are shown by the solid lines in Fig.\
\ref{f:omvb}. These are in good agreement with, and orthogonal to,
those of Lahav \etal\ (2002).

\subsection{Luminosity effects}

We note that the above analysis is independent of luminosity as we
examine everything at the effective luminosity of the survey,
$L_s$. From the correlation functions in different volume-limited
samples of 2dFGRS galaxies, Norberg \etal~(2001) found a luminosity
dependence of clustering of the form (cf. Eqn.\ \ref{e:r0star}),
\begin{equation}
b/b^{\ast} = 0.85 + 0.15(L/L^{\ast})
\end{equation}
which gives an estimate for the bias of the survey galaxies, $b_s =
1.06 b^{\ast}$ (using $L = 1.4L^{\ast}$), where $b^{\ast}$ is the bias
of $L^{\ast}$ galaxies. If this bias relation holds on the scales
considered in this paper then $\beta$ will be increased by the same
factor of 1.06, 
\begin{equation}
\beta(L^{\ast}, z_s) = 0.52 \pm 0.09
\end{equation}
and evolving $\beta$ in a `constant galaxy clustering' model (Lahav
\etal~2002) then,
\begin{equation}
\beta(L^{\ast}, z = 0) = 0.47 \pm 0.08,
\end{equation}
which we choose as a fiducial point to allow comparisons with other
surveys with different effective luminosities and redshifts. Lahav
\etal~(2002) obtained, $\beta(L^{\ast}, z = 0) = 0.50 \pm 0.06$, in
their combined 2dFGRS and CMB analysis, completely consistent with the
our result.

\subsection{Comparisons}
\label{s:omcomp}

Percival \etal~(2002) combined the 2dFGRS power spectrum with the
pre-{\it WMAP} CMB data, assuming a flat cosmology and found $\om(z=0)
= 0.31 \pm 0.06$. These measurements of $\om$ are also consistent with
a different estimation from the 2dFGRS and CMB (Efstathiou \etal\ 2002)
and from combining the 2dFGRS with cosmic shear measurements (Brown
\etal\ 2003).

Also plotted in Figure\ \ref{f:omvb} is the recent result from the
analysis of the {\it WMAP} satellite data. Spergel \etal~(2003) found
$\om = 0.29 \pm 0.07$ using {\it WMAP} data alone, although there are
degeneracies with other parameters. It is clear that this is
completely consistent with the other plotted contours. Spergel
\etal~(2003) also found that the epoch of reionisation, $\tau = 0.17$,
which would reduce the value of $b$ found by Lahav \etal\ (2002) by
about $16\%$, still in good agreement with the results in this paper.

\section{Summary}

In this paper we have measured the correlation function, and various
related quantities using 2dFGRS galaxies. Our main results are
summarised as follows:

\renewcommand{\theenumi}{(\roman{enumi})}
\renewcommand{\labelenumi}{\theenumi}

\begin{enumerate}
\item{The spherical average of $\xisp$ gives the redshift-space
correlation function, $\xis$, from which we measure the redshift space
clustering length, $s_0 = \as0\hmpc$. At large and small scales $\xis$
drops below a power law as expected, for instance, in the $\lcdm$
model.}

\item{The projection of $\xisp$ along the $\pi$ axis gives an estimate
of the real-space correlation function, $\xir$, which on scales $0.1 <
r < 12\hmpc$ can be fit by a power-law $(r/r_0)^{-\gamma_r}$ with $r_0
= \ar0\hmpc$, $\gamma_r = \agamr$. At large scales, $\xir$ drops below
a power-law as expected, for instance, in the $\lcdm$ model.}

\item{The ratio of real and redshift-space correlation functions on
scales of $8 - 30\hmpc$ reflects systematic infall velocities and
leads to an estimate of $\beta = \betaxirat$. The quadrupole moment of
$\xisp$ on large scales gives $\beta = \betaQ$.}

\item{Comparing the projections of $\xisp$ along the $\pi$ and
$\sigma$ axes gives an estimate of the distribution of random pairwise
peculiar velocities, $f(v)$. We find that large-scale infall
velocities affect the measurement of the distribution significantly
and cannot be neglected. Using $\beta=0.49$, we find that $f(v)$ is
well fit by an exponential with pairwise velocity dispersion, $a = 570
\pm 25\kms$, at small $\sigma$.}

\item{A multi-parameter fit to $\xisp$ simultaneously constrains the
shape and amplitude of $\xir$ and both the velocity distortion effects
parameterized by $\beta$ and $a$. We find $\beta = \betaoverall$ and
$a = \aoverall\kms$, using the Hubble Volume $\xir$ as input to the
model. These results apply to galaxies with effective luminosity, $L
\approx 1.4L^{\ast}$ and at an effective redshift, $z_s \approx
0.15$. We also find that the best fit values of $\beta$ and $a$ are
strongly correlated.}

\item{We evolve our value for the infall parameter to the present day
and critical luminosity and find $\beta(L=L^*, z = 0) = 0.47 \pm
0.08$. Our derived constraints on $\om$ and $b$ are consistent with a
range of other recent analyses.}

\end{enumerate} 

Our results show that the clustering of 2dFGRS galaxies as a whole is
well matched by a low density $\lcdm$ simulation with a non-linear
local bias scheme based on the smoothed dark-matter density
field. Nevertheless, there are features of the galaxy distribution
which require more sophisticated models, for example the distribution
of pairwise velocities and the dependence of galaxy clustering on
luminosity or spectral type. The methods presented have also been used
on sub-samples of the 2dFGRS, split by their spectral type (Madgwick
\etal\ 2003).

\section*{Acknowledgements}

The 2dF Galaxy Redshift Survey has been made possible through the
dedicated efforts of the staff of the Anglo-Australian Observatory,
both in creating the 2dF instrument and in supporting it on the
telescope. We thank Nelson Padilla, Andrew Benson, Sarah Bridle,
Yipeng Jing, Xiaohu Yang and Robert Smith for providing their results
in electronic form. We also thank the referee for valuable comments.

\section*{References}

\refer Baugh, C.M., Efstathiou, G., 1993, MNRAS, 265, 145

\refer Benson, A.J., Baugh, C.M., Cole, S., Frenk, C.S., Lacey, C.G.,
2000, MNRAS, 316, 107

\refer Brown, M.L., Taylor, A.N., Bacon, D.J., Gray, M.E., Dye, S.,
Meisenheimer, K., Wolf, C., 2003, MNRAS, 341, 100

\refer Cen, R., Bahcall, N.A., Gramann, M., 1994, ApJ, 437, 51

\refer Cole, S., Hatton, S., Weinberg, D.H., Frenk, C.S., 1998, MNRAS,
300, 945

\refer Colless, M. \& the 2dFGRS team, 2001, MNRAS, 328, 1039

\refer Davis, M., Peebles, P.J.E., 1983, ApJ, 208, 13

\refer Davis, M., Efstathiou, G., Frenk, C.S., White, S.D.M., 1985,
ApJ, 292, 371

\refer Diaferio, A., Geller, M., 1996, ApJ, 467, 19

\refer Efstathiou, G., 1988, in Proc. 3rd {\it IRAS} Conf., Comets to
Cosmology, ed. A.  Lawrence (New York, Springer), p. 312

\refer Efstathiou, G. \& the 2dFGRS team, 2002, MNRAS, 330, 29

\refer Evrard A.E. \etal, 2002, ApJ, 573, 7

\refer Guzzo, L. \etal, 2000, A\&A, 355, 1

\refer Hamilton, A.J.S., 1992, ApJ, 385, L5

\refer Hamilton, A.J.S., 1993, ApJ, 417, 19

\refer Hawkins, E., Maddox, S., Branchini, E., Saunders, W., 2001,
MNRAS, 325, 589

\refer Jenkins, A. \etal, 1998, ApJ, 499, 20

\refer Jing, Y.P., Mo, H.J., B\"orner, G., 1998, AJ, 494, 1

\refer Kaiser, N., 1987, MNRAS, 227, 1

\refer Kauffmann, G., Colberg, J.M., Diaferio, A., White, S.D.M.,
1999, MNRAS, 303, 188.

\refer Lahav, O., Lilje, P.B., Primack, J.R., Rees, M.J., 1991, MNRAS,
251, 128
 
\refer Lahav, O. \& the 2dFGRS team, 2002, MNRAS, 333, 961

\refer Landy, S.D., Szalay, A.S., 1993, ApJ, 388, 310

\refer Landy, S.D., Szalay, A.S., Broadhurst, T., 1998, ApJ, 494, L133
(LSB98)

\refer Landy, S.D., 2002, ApJ, 567, L1 (L02)

\refer Lewis, I.J. \etal, 2002, MNRAS, 333, 279

\refer Limber, D.N., 1954, ApJ, 119, 655

\refer Lin, H., Kirshner, R.P., Shectman, S.A., Landy, S.D., Oemler, A.,
 Tucker, D.L., Schechter, P.L., 1996, ApJ, 464, 60

\refer Loveday, J., Efstathiou, G., Peterson, B.A., Maddox, S.J.,
1992, ApJ, 400, 43

\refer Loveday, J., Maddox, S.J., Efstathiou, G., Peterson, B.A.,
1995, ApJ, 442, 457

\refer Maddox, S.J., Efstathiou, G., Sutherland, W.J., 1990, MNRAS,
246, 433

\refer Maddox, S.J., Efstathiou, G., Sutherland, W.J., 1996, MNRAS,
283, 1227

\refer Madgwick, D.S. \& the 2dFGRS team, 2003, MNRAS, submitted, astro-ph/0303668

\refer Magliochetti, M., Porciani, C., 2003, MNRAS, submitted,
astro-ph/0304003

\refer Marzke, R.O., Geller, M.J., da Costa, L.N., Huchra, J.P., 1995,
AJ, 110, 477

\refer Norberg, P. \& the 2dFGRS team, 2001, MNRAS, 328, 64

\refer Norberg, P. \& the 2dFGRS team, 2002a, MNRAS 336, 907

\refer Norberg, P. \& the 2dFGRS team, 2002b, MNRAS, 332, 827

\refer Padilla, N., Baugh, C.M., 2003, MNRAS, submitted,
astro-ph/0301083

\refer Peacock, J.A., Smith, R.E., 2000, MNRAS, 318, 1144

\refer Peacock, J.A. \& the 2dFGRS team, 2001, Nature, 410, 169 

\refer Peebles, P.J.E., 1980, {\it The Large Scale Structure of the
Universe}, Princeton Univ. Press, Princeton, NJ

\refer Percival, W. \& the 2dFGRS team, 2001, MNRAS, 327, 1297

\refer Percival, W. \& the 2dFGRS team, 2002, MNRAS, 337, 1068

\refer Press, W., Teukolsky, S., Vetterling, W., Flannery, B., 1992,
{\it Numerical Recipes}, Cambridge University Press, Cambridge

\refer Ratcliffe, A., Shanks, T., Parker, Q.A., Fong, R., 1998, MNRAS,
296, 191

\refer Saunders, W., Rowan-Robinson, M., Lawrence, A., 1992, MNRAS,
258, 134 (S92)

\refer Sheth, R.K., 1996, MNRAS, 279, 1310

\refer Spergel, D.N. \etal\ (The {\it WMAP} Team), 2003, ApJ, submitted, astro-ph/0302209

\refer Tucker, D.L. \etal, 1997, MNRAS, 285, 5

\refer Verde, L. \& the 2dFGRS team, 2002, MNRAS, 335, 432

\refer Yang, X., Mo, H.J., Jing, Y.P., van den Bosch, F., Chu, Y,
2003, MNRAS, submitted, astro-ph/0303524

\refer Zehavi, I. \etal\ (The SDSS Collaboration), 2002, ApJ, 571, 172

\refer Zehavi, I. \etal\ (The SDSS Collaboration), 2003, ApJ, submitted, astro-ph/0301280

\appendix

\section[]{COHERENT INFALL EQUATIONS}
\label{s:a}

Kaiser (1987) pointed out that the coherent infall velocities take a
simple form in Fourier space,
\begin{equation}
P_s(k) = (1 + \beta\ct_k^2)^2 P_r(k).
\end{equation}
Hamilton (1992) completed the translation of these results into real
space,
\begin{equation}
\xi'(\sigma, \pi) = [1 + \beta(\partial/\partial z)^2(\nabla^2)^{-1}]^2 \xir,
\end{equation}
which reduces to
\begin{equation}
\xi'(\sigma, \pi) = \xi_0(s)P_0(\ct) + \xi_2(s)P_2(\ct) + \xi_4(s)P_4(\ct),
\end{equation}
where in general,
\begin{equation}
\xi_0(s) = \left(1 + \frac{2\beta}{3} + \frac{\beta^2}{5}\right)\xir,
\end{equation}
\begin{equation}
\xi_2(s) = \left(\frac{4\beta}{3} + \frac{4\beta^2}{7}\right)[\xir-\overline{\xi}(r)],
\end{equation}
\begin{equation}
\xi_4(s) = \frac{8\beta^2}{35}\left[\xir + \frac{5}{2}\overline{\xi}(r)
-\frac{7}{2}\overline{\overline{\xi}}(r)\right],
\end{equation}
and
\begin{equation}
\overline{\xi}(r) = \frac{3}{r^3}\int^r_0\xi(r')r'{^2}dr',
\end{equation}
\begin{equation}
\overline{\overline{\xi}}(r) = \frac{5}{r^5}\int^r_0\xi(r')r'{^4}dr'.
\end{equation}
In the case of a power law form for $\xir$ these equations reduce to
the form shown in Eqns.\ \ref{e:xi0} to \ref{e:xi4}. In the case of
non-power law forms for the real-space correlation function these
integrals must be performed numerically.

\bsp
\label{lastpage}
\end{document}